\documentclass[a4paper,12pt,useAMS]{mn2e}

\usepackage{graphicx}

\title[Improved Lens Model of Cl0024+1654]{New Multiply-Lensed Galaxies Identified in ACS/NIC3 Observations of Cl0024+1654 Using an Improved Mass Model}

\author[Adi Zitrin et al.]{Adi Zitrin$^{1}$\thanks{E-mail:
adiz@wise.tau.ac.il}, Tom Broadhurst$^{1}$, Keiichi Umetsu$^{2,3}$,
Dan Coe$^{5}$, Narciso Ben\'itez$^{4}$, \and Bego\~na Ascaso$^{6}$, Larry Bradley$^{5}$, Holland Ford$^{5}$, James Jee$^{6}$, Elinor Medezinski$^{1}$, \and Yoel Rephaeli$^{1}$, Wei Zheng$^{5}$\\\\
$^{1}$The School of Physics and Astronomy, the Raymond and Beverly Sackler Faculty of Exact Sciences, Tel Aviv University,\\ Tel Aviv 69978, Israel\\
$^{2}$Institute of Astronomy and Astrophysics, Academia Sinica, P.~O. Box 23-141, Taipei 10617, Taiwan\\
$^{3}$LeCosPA, National Taiwan University, Taipei 10617, Taiwan\\
$^{4}$Instituto de Astrof\'isica de Andaluc\'ia (CSIC), C/Camino Bajo de Hu\'etor, 24, Granada, 18008, Spain\\
$^{5}$Department of Physics and Astronomy, Johns Hopkins University, 3400 North Charles Street, Baltimore, MD 21218\\
$^{6}$Department of Physics, University of California, One Shields Avenue, Davis, CA 95616, USA}

\begin{document}

\pagerange{\pageref{firstpage}--\pageref{lastpage}} \pubyear{2009}

\maketitle

\label{firstpage}

\begin{abstract}

We present an improved strong-lensing analysis of Cl0024+1654 ($z$=0.39)
using deep HST/ACS/NIC3 images, based on 33 multiply-lensed images of
11 background galaxies. These are found with a model that assumes mass
approximately traces light, with a low order expansion to allow for
flexibility on large scales. The model is constrained initially by the
well known 5-image system ($z$=1.675) and refined as new
multiply-lensed systems are identified using the model.  Photometric
redshifts of these new systems are then used to constrain better the
mass profile by adopting the standard cosmological relation between
redshift and lensing distance. Our model requires only 6 free
parameters to describe well all positional and redshift data.  The
resulting inner mass profile has a slope of $d\log M/d\log r\simeq
-0.55$, consistent with new weak-lensing measurements where the data
overlap, at $r\simeq200$ kpc/$h_{70}$. The combined profile is well
fitted by a high concentration NFW mass profile, $C_{\rm vir}\sim
8.6\pm1.6$, similar to other well studied clusters, but larger than
predicted with standard $\Lambda$CDM. A well defined radial critical
curve is generated by the model and is clearly observed at $r \simeq
12\arcsec$, outlined by elongated images pointing towards the centre
of mass. The relative fluxes of the multiply-lensed images are found
to agree well with the modelled magnifications, providing an
independent consistency check.

\end{abstract}


\section[]{Introduction}

The rich cluster Cl0024+1654 is one of the most distant clusters
discovered by Zwicky (1959) and displays one of the finest examples of
gravitational lensing (see also Broadhurst et al.  2000). Many arcs and images of distant lensed sources
are visible, among them a 5-image system of a well-resolved galaxy,
which was first noted by Koo (1988, see also Smail et al. 1996). The
arc mentioned by Koo (1988) was later resolved into a close triplet of
arcs by Kassiola, Kovner, \& Blandford (1992), while marking two
additional images which were later found to be members of this system
by Smail et al. (1996) and by Colley, Tyson, \& Turner (1996), using
HST WFPC-1 and WFPC-2 data, respectively. These arcs have been used by
Colley, Tyson, \& Turner (1996) to construct an image of the source, and
by Tyson, Kochanski, \& dell'Antonio (1998) to examine details of the
mass distribution.  The redshift of this system, $z=1.675$ (Broadhurst
et al. 2000), was eventually obtained from weak interstellar
absorption features, permitting an accurate and fairly
model-independent mass for the central area of
M($<$100kpc/h)=$1.11\pm0.03\times 10^{14}h^{-1}$M$_{\odot}$ and a
mass-to-light ratio of
M/L$_B$($<$100kpc/h)=320h$\pm$30h(M/L$_B$)$_{\odot}$, because of the
symmetric arrangement of the images (Broadhurst et al.  2000).

Lens models for Cl0024+1654, and strong-lensing (SL) models in
particular, have generally improved with higher quality imaging. Using
ground based data, Kassiola, Kovner, \& Blandford (1992) and
Wallington, Kochanek, \& Koo (1995) reproduced fits to the close triplet of arcs
(A,B,C, in their notation; images 1.3, 1.4, 1.5 in Figure
\ref{fig:1_z} here) but considered arc D (image 1.1 in Figure
\ref{fig:1_z} here) an unlikely counter image. Subsequent HST images
revealed that A,B,C and D are morphologically similar in detail (Smail
et al. 1996) and that a further radially directed arc in the cluster
centre, E (image 1.2 in Figure \ref{fig:1_z} here), is another
complete image of the same source (Colley, Tyson, \& Turner 1996, see also Broadhurst et al.  2000). A
512 parameter fit to the resolved imaging data for the arc system was presented by Tyson, Kochanski, \& dell'Antonio (1998). This
solution required the inclusion of a number of small dark deflecting
`mascons' around each of the lensed images to offset the symmetry of a
dominant central potential (see resulting mass distribution in Fig. 2
of Tyson, Kochanski, \& dell'Antonio 1998, see also Broadhurst et al.  2000) for which no corresponding
cluster members are visible.

A simpler model of this cluster mass-distribution was presented by
Broadhurst et al. (2000), by assigning profiles to the brightest
cluster members using the form advocated by NFW (Navarro, Frenk \&
White 1997) in a 16 parameter fit, which reproduced very accurately
the 5-image system. This model identified another lensed system
comprising two multiple-images, which have been since seen to be of
very similar colours and morphology, also in recent deep ACS
images. This system is used by Jee et al. (2007) together with the
weak-lensing (WL) distortions (Object B1,2 in the notation of Jee et
al. 2007) to derive the inner mass distribution based on a
non-parametric technique (e.g., Abdelsalam, Saha, \& Williams 1998, Diego
et al. 2005). Generally, ringlike or monopole degeneracies in such techniques may limit the plausibility of
obtaining a reliable solution when only SL data are used to constrain the model (Coe et al. 2008, Liesenborgs et al. 2008a,b). Note that here we
found a new counter image of this system on the inner side of the
radial critical curve, which was also recently predicted by the model of Liesenborgs et al. (2008b). This is explained in the Results section and is
seen Figures \ref{fig:2_z} and \ref{fig:2_sub}.

Strong-lensing modelling methods have improved over the past two decades,
in response to computational advances and higher quality data, yet most of these methods still involve
many parameters. Most methods can be classified as ``parametric" if
based on physical parameterisation, and as ``non-parametric" if they
are ``grid-based" (see also \S 4.4 in Coe et al. 2008, and references
therein). Currently, both methods include too many parameters to be
well-constrained by the number of initially known multiply-lensed
systems. Here we use the deep ACS imaging to identify new
multiply-lensed systems, motivated by the successful approach of
Broadhurst et al. (2005a) for identifying new multiply-lensed systems
with a minimalistic approach to the lens modelling. We present an improved
modelling method which involves only 6 free parameters, enabling
easier constraint by known systems, since the number of constraints
has to be equal or larger to the number of parameters in order to get
a reliable fit. Two of these parameters are primarily set to reasonable values and so only 4 of these parameters have to be constrained initially, which sets a very reliable starting-point using the known 5-image system. This we find is sufficient for finding many multiple-images, which are then iteratively incorporated into the model, by using their photo-$z$ estimations to constrain the two initially-set parameters which control the mass profile slope. This will be explained in more detail in \S \ref{model}. Our modelling relies on the reasonable expectation that the mass distribution
approximately traces the galaxy distribution and is smoothly varying
so we may prefer solutions in which the large scale mass distribution
has a minimum of structure. The individual contribution to lensing
from the visible galaxies must be included which adds a small well-defined contribution to the overall deflection field. We also compare
our resulting mass profile with WL measurements made from deep
multi-colour Subaru imaging (Umetsu et al. 2009, in prep) to examine
the consistency of the model in the region of overlap.

A major motivation for pursuing accurate lensing-maps is the increased
precision of model predictions for cluster-size massive halos in the
standard $\Lambda$CDM model for the formation (see Umetsu \& Broadhurst 2008, e.g., Bullock et al.
2001, Hennawi et al. 2007, Neto et al. 2007, Duffy et al. 2008). Many of the
free parameters of this model now rest on a firm empirical foundation
with relatively tight constraints on the index and normalisation of
the power spectrum of density perturbation and the background
cosmological model (see Umetsu \& Broadhurst 2008, e.g., Spergel et al. 2003, Tegmark et al. 2004,
Spergel et al. 2007).

The standard $\Lambda$CDM model is amenable to comparisons with the real Universe via advanced $N$-body simulations, in particular the recent Millennium simulation (Springel et al. 2005) which simulates a huge volume of $500 {\rm
Mpc}/h$, and has been used to predict the mass function and evolution
of nearly 100,000 group and cluster-size CDM halos. Clusters of galaxies and the effects of gravitational potential in them (such as lensing), are good candidates for such comparisons, since there, baryons which
are usually omitted from dynamical N-body simulations, are presumed not to have a significant impact on the shape of the cluster gravitational potential (e.g., Blumenthal 1986, Broadhurst \& Barkana 2008, Umetsu \& Broadhurst 2008). This is because the high
temperature of the cluster gas prevents efficient cooling and hence
the majority of baryons simply trace the gravitational potential of
the dominant dark matter (see also Umetsu \& Broadhurst 2008). Massive clusters are of particular interest
in the context of this model, because they are predicted to have a
distinctively shallow mass profile (or low concentration) described by the
form proposed by Navarro, Frenk, \& White (1997).

In earlier papers we have explored the combination of weak and strong
lensing with new methods designed to achieve the maximum possible
lensing precision, combining all lensing information (Broadhurst et
al. 2005a,b, Medezinski et al. 2007,
Broadhurst et al. 2008, Umetsu \& Broadhurst 2008), applied initially to A1689 and then expanding
to several other massive well-known clusters with similarly high
quality data. In Broadhurst et al. (2005b) we developed a
model-independent method (see also Umetsu \& Broadhurst 2008) for reconstructing the cluster mass profile
using azimuthally-averaged WL shape distortion and magnification bias
measurements, in the wide-field Subaru images. This together with many
multiple images identified in deep {\it Hubble Space Telescope} (HST)
Advanced Camera for Surveys (ACS) imaging, defined a detailed
lensing-based cluster mass profile, out to the cluster virial radius
($r \sim 2 h^{-1}$ Mpc). The combined strong and weak lensing mass
profile has been well fitted by an NFW profile (Navarro, Frenk, \& White
1997) with high concentration of $c_{\rm vir}\sim 13.7$. This value is
significantly larger than theoretically expected ($c_{\rm vir}\simeq
5$) for the standard $\Lambda$CDM model (Bullock et al.
2001, Neto et al. 2007, Duffy et al. 2008), even after accounting for expected projection
bias (Oguri et al. 2005, Hennawi et al. 2007, Corless, King \& Clowe
2008). While this discrepancy is weakened when various inherent modelling and simulation uncertainties are taken into account (Sadeh \& Rephaeli 2008), it still raises serious questions regarding the basic assumptions behind the $\Lambda$CDM model.

This tendency for higher concentrations than expected by the $\Lambda$CDM model, is confirmed also in our recent
lensing measurements of four other well-known clusters which all have
WL profiles very similar to A1689, and for which the concentrations lie
in the range $8<c_{\rm vir}<12$ (Broadhurst et al.  2008). In addition,
independent information on the internal dynamics of A1689 from over
1000 galaxy spectra (Lemze et al. 2008b) and deep X-ray data (Lemze et
al. 2008a) independently support the high concentration claimed for
A1689. Other work is also establishing high concentration profiles as
this norm for galaxy clusters, which includes other well studied
clusters (Gavazzi et al. 2003, Kneib et al. 2003, Oguri et al. 2009), samples of X-ray
selected clusters (Duffy et al. 2008), and clusters with large
Einstein radii (Broadhurst \& Barkana 2008). This is very significant since similarly large Einstein radius (and high concentration-parameter) clusters, would require an earlier formation of the large scale structure than implied by the standard $\Lambda$CDM model (Sadeh \& Rephaeli 2008).

The paper is organised as follows: In \S 2 we describe the
observations. In \S 3 we describe the photometry procedure and its
processing for obtaining photometric redshifts. In \S 4 we detail the
modelling method and its implementation. We characterize the
model, and explain how it was constrained and the verification
criteria it obeys. In \S 5 we report and discuss the results,
particularly the newly-discovered multiply lensed systems and the mass
distribution. In \S6 we summarize and conclude this work.

Throughout this paper, we use the AB magnitude system, and adopt a concordance
$\Lambda$CDM cosmology with ($\Omega_{\rm m0}=0.3$, $\Omega_{\Lambda
0}=0.7$, $h=0.7$). With these parameters one arcsecond corresponds corresponds to the
physical scale of 5.3 kpc$/h_{70}$ for this cluster. The reference centre
of our analysis is fixed at the centre of the cD galaxy:
RA = 00:26:35.7, Dec = +17:09:43.1 (J2000.0).

\section[]{Observations and target selection}

Observations of this cluster were performed in the framework of the
ACS Guaranteed Time Observations (GTO) which includes deep
observations of several massive, intermediate-redshift galaxy
clusters. As mentioned in Broadhurst et al.\ (2005a), some important
aims of the GTO program are determination of the mass distribution of
clusters for testing the standard cosmological model and to study
distant, background lensed galaxies for which some of the very highest
redshift galaxies are known because of high magnification by massive
clusters (Franx et al. 1997, Frye \& Broadhurst 1998, Frye ,
Broadhurst \& Ben\'itez 2002, Kneib et al. 2004, Stark et al. 2007,
Bouwens et al. 2008, Bradley et al. 2008, Zheng et al. 2009).

Although the SL by Cl0024+1654 has been analysed before, only two
multiple-image systems are currently known in this cluster, including
the classic 5-image system (Smail et al. 1996, Colley, Tyson \&
Turner 1996) and a pair of fainter images identified in WFPC-2 data by
the model of Broadhurst et al. (2000). One more multiple system is
claimed by Jee et al. (2007), which is discussed later in \S 5.

The relatively large Einstein radius of this cluster of $\sim
30\arcsec$ at a source redshift of $z_s=1.675$ (corresponding to the 5
image system, Broadhurst et al. 2000) encourages us to search for more
such multiple images in our deep GTO imaging. For comparison, in
similar quality GTO imaging of A1689, 33 multiply-lensed
galaxies were uncovered, forming 106 multiple images for which the Einstein radius is
larger, $\sim 45\arcsec$ ($z\sim 1.5$). Therefore, taking simply the ratio of
the Einstein radius squared, we may expect to detect approximately
$\sim 13$ lensed sources, and where the corresponding number of
multiple images depends somewhat on the degree of substructure within
the Einstein radius.

Cl0024+1654 was observed in November 2004, with the Wide Field Channel
(WFC) of the ACS installed on HST. Integration times of 6435, 5072,
5072, 8971, 10144, and 16328 seconds, were obtained through the F435W,
F475W, F555W, F625W, F775W, and F850LP filters, respectively. The
NICMOS/NIC3 images of Cl0024+1654 were obtained in 2007 July, with
exposures of 9706 seconds in both the F110W and F160W bands. We also
used additional archival NICMOS images of this field, taken in 2006
August. The NICMOS images were processed with a custom pipeline
(partially based on IRAF scripts) written at the University of
California Santa Cruz ("NICRED"; Magee, Bouwens, \& Illingworth,
2007).

\section[]{Photometry and photometric redshifts}

We obtain $B V g\arcmin r\arcmin i\arcmin z\arcmin J H$ photometry
from HST ACS and NIC3 (NICMOS C3) images.  The ACS images were
initially reduced, processed, and analysed by APSIS, the ACS GTO
pipeline (Blakeslee et al. 2003).  An optimal $\chi^2$ detection image
was created as a weighted sum of all filters, each divided by its
background RMS.  Objects were detected and photometry obtained using
SExtractor (Bertin \& Arnouts 1996).

SExtractor does well to detect most of our multiple image candidates,
but some of the fainter objects elude detection with our choice of
parameters.  For those objects, we construct apertures ``by hand'' and
use SExSeg (Coe et al. 2006, included in the ColorPro package) to
force SExtractor to obtain photometry for those objects.

Other objects are lost in the glare of nearby cluster members.  To
uncover these objects, we carefully modelled and subtracted the
light from most of the cluster galaxies in each of the six filters.
First of all, we subtracted the central cD galaxy and its halo from the cluster
centre. This was achieved by first masking all the galaxies in the frame, except for the
cD galaxy, using SExtractor, and masking manually the small satellite galaxies with an IRAF
task built for this purpose, which adopts the shape of the galaxies for the careful masking
of their light. Then, we fitted an elliptical model to the cD galaxy with the IRAF
tasks ELLIPSE and BMODEL. Afterwards, we subtracted the model from the image and the
resultant image was used to estimate and subtract the background and the cD halo
with SExtractor. This procedure is iterated until the halo is completely subtracted.
Usually, the procedure converges in two iterations (Ascaso et al., in prep.). After
that, the surrounding galaxies were fitted into two components by a S\'ersic plus
exponential profile and therefore their model was again subtracted. The small
residua in the final image were also cleaned with SExtractor.
This allows us to detect new images (e.g., Figure \ref{fig:2_sub}), including faint central
demagnified images.

This galaxy-subtraction procedure should improve the photometry (and thus
photo-$z$ estimation) for those images close to cluster galaxies. However,
the subtraction tends to adversely affect the colours of nearby
objects, due to the fact that the
best fit model of the cluster galaxies changes its orientation in different bands so it is difficult to model the wings of cluster galaxies in
all filters sufficiently well to avoid this.

The ColorPro software is used to ensure robust colours across the ACS
and NIC3 images. The NIC3 images are co-registered to the ACS
coordinates. Identical photometric apertures are applied to all the
aligned images. And corrections are made for broader NIC3 PSF.

Given our $B V g\arcmin r\arcmin i\arcmin z\arcmin J H$ photometry, we
obtain photometric redshifts using BPZ v1.99.2 (Ben\'itez 2000,
Ben\'itez et al. 2004, Coe et al. 2006).  The distances to the
galaxies are, of course, key ingredients to the lens model.  The BPZ
analysis also aids us in our multiple image identification, as
multiple images should all have the same SEDs (spectral energy
distributions).

In Figure \ref{fig:seds} we compare the SEDs of the multiple image
candidates of each lensed galaxy.  We expect good agreement when the
images are well isolated, as is the case for images 1.1, 1.3, and 1.4,
for example.  But some images, such as 1.2 and 1.5, are contaminated
by the light from nearby cluster galaxies. We conclude that the photometric redshift estimation for these contaminated galaxies is less reliable, even in the galaxy-subtracted image, as explained above. We can tell which are
likely to be contaminated by a visual inspection of the images.

\section[]{Strong Lensing Model}\label{model}

Our aim is to develop a SL modelling method with a minimum of
free parameters, so that we have the predictive power to find new
multiply-lensed systems. The basic assumption in the construction of the
model is that the observed galaxy distribution approximately follows the
general DM distribution of the cluster. We assign a fixed power-law
profile to cluster member galaxies, of slope $q$, and scaled linearly in amplitude by
the observed brightness. The power law index $q$ is a free parameter of the model. Adding all these profiles together results in a fairly smooth
overall mass distribution with local maxima corresponding to each member galaxy.
This approach is similar to that of Broadhurst et al.\ (2005a), which was successfully used to identify
over 30 sets of multiple images in similarly deep ACS/GTO imaging of
A1689, and also for the more distant cluster SDSS1004+4112 which was discovered to have 4 bright
lensed QSO images (Oguri et al. 2004, Inada et al. 2005) and for which 3
sets of new multiply lensed galaxies were identified in deep ACS
images (Sharon et al. 2005).

We expect that the DM distribution is smoother than this co-added
galaxy distribution and with a much higher overall mass scaling. To
represent the DM distribution we interpolate over the above sum of galaxy masses
with a low order cubic-spline interpolation and calculate
the corresponding deflection field, where the polynomial order of this function is denoted as
$S$, or the $smoothing~degree$, which is a free parameter of the model. The scaling of this smooth DM component
relative to the total galaxy mass is denoted by $K_{gal}$, which is another
free parameter of the model.

Though the best fitting solutions require a relatively small galaxy
contribution compared to the smooth component, the galaxy contribution
is, however, not negligible in the sense that lensed images lying near
cluster members are locally deflected significantly by them, and hence
these must be included in order to correctly identify multiply-lensed
images. We denote an additional free parameter of the model as
$K_{q}$, which as seen in eqs. \ref{deflection} and \ref{deflection2},
is proportional to the lensing distance ratio $d_{ls}/d_{s}$, and
contains the other constants seen there.

We do not expect the galaxies and DM to trace each other in detail and some flexibility should be allowed. Therefore we simply Taylor
expand to first order the potential of the DM distribution, which
adds two more free parameters: the shear
amplitude, $\gamma$, and its position
angle $\phi$, which together equivalently describe the overall matter ellipticity.

We find that these 6 parameters are sufficient for identifying new multiple images and for defining well the mass profile of the
cluster as we show in detail below. This is preferable to the common approach of subjectively defining sub-clusters as separate elliptical masses, with the many attendant extra parameters this entails.

In the process of identifying new images it is important to make use
of all the pixel information, and so we delens each set of pixels belonging to an image
in each passband back to the source plane to act as the source for generating
counter images. These relensed images then reflect the internal colour
and morphological structure of the observed lensed candidate galaxy,
which is very helpful in uniquely identifying new images. We generate
a family of relensed images to cover a range of plausible lensing
distances as any image must lie somewhere along a locus defined
this way. We now detail the modelling procedure.

\subsection[]{Initial Mass Distribution}

A catalogue of the cluster galaxies was created by colour-colour
diagrams. The $\sim300$ brightest cluster galaxies (within the ACS
frame) were chosen in order to construct the mass distribution. The
list includes the centre pixel coordinates of each galaxy and its
flux. By assuming a certain $M/L$ ratio, or that the flux is
proportional to the mass, the deflection field contributed by each
object can now be calculated by assigning a galaxy surface-density
profile for each galaxy, $\Sigma(r)=Kr^{-q}$, which is integrated to
give the interior mass, $M(<\theta)=\frac{2\pi
K}{2-q}(d_{l}\theta)^{2-q}$. This results in a deflection angle of
(due to a single galaxy):
\begin{equation}
\label{deflection}
 \alpha(\theta)= \frac{4GM(<\theta)}{c^2\theta}\frac{d_{ls}}{d_{s}d_{l}}.
\end{equation}
Since $M(<\theta)\propto\theta^{2-q}$, and since we assume a certain
$M/L$ relation according to which the flux, F, is proportional to the mass, we can reduce the latter formula to
get:
\begin{equation} \label{deflection2}
 \alpha(\theta)= K_{q}F\theta^{1-q} ,
\end{equation}
where $K_{q}$ is a new defined constant which contains all previous
constants and the proportion relations, and it is dependent also upon the power-law index,
$q$.

The deflection angle in a certain point $\vec{\theta}$
due to lumpy galaxy components is
simply a linear
sum of each galaxy
contribution, summed over all $i$ galaxies:
\begin{equation}
\label{deflection3}
 \vec{\alpha}_{gal}(\vec{\theta})= K_{q}\sum_{i}
F_i\, |\vec{\theta}-\vec{\theta}_i|^{1-q}
\frac{\vec{\theta}-\vec{\theta}_i}{|\vec{\theta}-\vec{\theta}_i|}.
\end{equation}

A discretised version of equation (\ref{deflection3}) over a two
dimensional square grid $\vec\theta_m$ of $N\times N$ pixels is given by:
\begin{eqnarray}
\label{deflection_x1}
 \alpha_{gal,x}(\vec\theta_m)=K_{q}\sum_{i}
F_i\, \left[
 (\Delta x_{mi})^2
 +
 (\Delta y_{mi})^2
\right]^{-q/2}
\Delta x_{mi},\\
\label{deflection_y1}
 \alpha_{gal,y}(\vec\theta_m)=K_{q}\sum_{i}
F_i\, \left[
 (\Delta x_{mi})^2
 +
 (\Delta y_{mi})^2
\right]^{-q/2}
\Delta y_{mi},
 \end{eqnarray}
where $(\Delta x_{mi},\Delta y_{mi})$ is the displacement vector
$\vec\theta_m-\vec\theta_i$ of the $m$th pixel point, with respect to the
$i$th galaxy position $\vec\theta_i$.

From this a deflection field for the galaxy contribution is easily
calculated analytically as above, and the mass distribution is now
rapidly calculated locally from the divergence of the deflection field,
i.e., the 2D equivalent of Poisson's equation.

\subsection[]{The Dark Matter Distribution}\label{ss:DM}

The mass contribution of galaxies, described above, is anticipated to
comprise only a small fraction of the total mass of the cluster, which
is expected to be dominated by a smooth distribution of DM. We now
simply assume that the galaxies approximately trace the dark
matter. This assumption was found to work well in earlier work on this
cluster and on A1689 (Broadhurst et al. 2000, Broadhurst et al. 2005a)
where an unprecedented number of multiple-images were found (Broadhurst
et al. 2000, Broadhurst et al. 2005a). Since the DM is of course
expected to be smoother than the distribution of galaxies, we smooth
the initial guess of the DM distribution obtained above, choosing for
convenience a low-order cubic spline interpolation.

The smoothing degree (the polynomial degree, $S$) is also a free
parameter of the model. The deflection field of the DM is then (where each pixel is a $\delta$-function mass distribution):
\begin{equation}
\label{deflection_xDM}
 \alpha_{DM,x}(\vec\theta_m)=
K_{q}\sum_{i} P_i\, \left[
 (\Delta x_{mi})^2
 +
 (\Delta y_{mi})^2
\right]^{-1}
\Delta x_{mi},
 \end{equation}
 \begin{equation}
\label{deflection_yDM}
 \alpha_{DM,y}(\vec\theta_m)=
K_{q}\sum_{i} P_i\, \left[
 (\Delta x_{mi})^2
 +
 (\Delta y_{mi})^2
\right]^{-1}
\Delta y_{mi},
 \end{equation}

where $P_{i}$ represents the mass value in the $i$th pixel. Thus we
obtain now the deflection field due to the DM, hereafter
$\vec\alpha_{DM}(\vec\theta)$,
or the $\it{smooth~component}$.

\subsection[]{The Initial Deflection Field}

After obtaining the two components of the deflection field, we now
simply add them together to get a total deflection field as follows:
\begin{equation}
\label{defTot}
\vec{\alpha}_T(\vec\theta)=K_{gal} \vec\alpha_{gal}(\vec\theta)+(1-K_{gal})\vec\alpha_{DM}(\vec\theta),
\end{equation}

where $K_{gal}$ is the relative contribution of the galaxy component
to the deflection field.

\subsection[]{External Shear}

Since the assumption that the DM follows the galaxies light or mass is
not expected to be rigorous in detail, allowance should be made for
the unknown DM distribution to differ somewhat from the light,
particularly on large scales where the influence of mass outside the
central region may be important through tidal interaction. Various ways
have been used before, among them matching a high-order polynomial
which would fine-tune the DM distribution (e.g., Broadhurst et
al. 2000, 2005a). However, we find that adding an external shear, or
equivalently a large scale ellipticity (e.g., Kovenr 1987a,b, Keeton,
Kochanek, \& Seljak, 1997, and references therein) provides
considerable flexibility and adds only two additional free parameters,
if we only Taylor expand the potential to first-order. Note that this
is also ``safer" than other methods since it relies on physical considerations
alone. Thus, the total deflection field is now given by:

\begin{equation}
\label{defTotAdd}
\vec\alpha_T(\vec\theta)= K_{gal} \vec{\alpha}_{gal}(\vec\theta)
+(1-K_{gal}) \vec\alpha_{DM}(\vec\theta)
+\vec\alpha_{ex}(\vec\theta),
\end{equation}
where the deflection field at position $\vec\theta_m$
due to the external shear,
$\vec{\alpha}_{ex}(\vec\theta_m)=(\alpha_{ex,x},\alpha_{ex,y})$,
is given by (see also Kovenr 1987a,b,
Keeton, Kochanek, \& Seljak, 1997, and references therein):
\begin{equation}
\label{shearsx}
\alpha_{ex,x}(\vec\theta_m)
= |\gamma| \cos(2\phi_{\gamma})\Delta x_m
+ |\gamma| \sin(2\phi_{\gamma})\Delta y_m,
\end{equation}
\begin{equation}
\label{shearsy}
\alpha_{ex,y}(\vec\theta_m)
= |\gamma| \sin(2\phi_{\gamma})\Delta x_m -
  |\gamma| \cos(2\phi_{\gamma})\Delta y_m,
\end{equation}
where $(\Delta x_m,\Delta y_m)$ is the displacement vector of the
position $\vec\theta_m$ with respect to a fiducial reference position,
which we take as the lower-left pixel position $(1,1)$, and
$\phi_{\gamma}$ is the position angle of the spin-2 external
gravitational shear measured anti-clockwise from the $x$-axis.

\subsection[]{The Magnification and Critical Curves}

The magnification can be calculated simply from the gradients of the
above deflection field. The magnification at a given position (see
Narayan, \& Bartelmann 1996, Broadhurst et al.\ 2005a, and references
therein), $\mu(\vec\theta)$,
is given by the Jacobian of the lens
mapping and can be expressed by the derivatives of the deflection field:
\begin{equation}
\centering{
\label{magnification}
\mu(\vec\theta)^{-1}=1-\vec\nabla\cdot \vec{\alpha}+\alpha_{x,x}\alpha_{y,y}-\alpha_{x,y}\alpha_{y,x}},
\end{equation}

where
$\vec\alpha(\vec\theta)$
 and its derivatives refer to the total deflection field (now containing
 also the external shears),
$\vec\alpha_T(\vec\theta)$, and its
 derivatives, respectively. It should be noted that
$\alpha_{x,y}=\alpha_{y,x}$
because the deflection field is curl free. After constraining the model parameters, we calculate
the magnification field using the above equation, which depends
on the source distance, via the lensing distance ratio, $d_{ls}/d_s$.
The relative magnifications are used later as a consistency check of our modelling,
by comparison with the independent relative flux information for the multiply-lensed images.

\subsection[]{Iteration process}
We have found that the search for an adequate SL model can be divided
conveniently into two stages of iteration. The first stage (hereafter
the $internal~iteration$), is to constrain the parameters -
$K_{q}$, $K_{gal}$, $\gamma$, and $\phi_\gamma$. The slope
determining parameters, $q$ and $S$, are first held fixed at reasonable
values, since these parameters are not strongly constrainable at this
point, until we have identified new lensed systems and
incorporated their photo-$z$ measurements to break the mass-sheet
degeneracy and thereby obtain a useful measure of the gradient of the
mass profile.
We start by minimising as above from de-lensing only the 5-image system, fully constraining these 4 internal parameters. For this internal iteration, we find a source plane iteration is sufficient to begin our search for new multiply-lensed systems. This minimisation is efficient and allows
us to search exhaustively in an interactive way for new
multiply-lensed systems, using each pixel information as described above. The
properties of the new multiply-lensed systems, including photo-$z$
estimations, are outlined in the Results section and listed in Table
\ref{systems}.

Having obtained new multiple images this way we have discovered that
we can make use of their photo-$z$ information to meaningfully constrain
the slope of the mass profile. We generate a grid of values covering a
wide range of $q$ and $S$. We now obtain the best fit by minimising
the model solution per each predetermined combination of these parameters, incorporating the photo-$z$ estimations of these
new systems. This we name the $outer~iteration$. We maintain the
minimisation in the image-plane in order to avoid the inherent bias
towards shallower mass profiles when relying on source-plane
minimisations. We also check that all the observed images are
reproduced in this process and also examine for consistency the
relative image brightness compared with the model-predicted relative
magnifications. In Figures \ref{fig:mass_comp} to \ref{fig:mag_compS} we show the effects of different $q$ and $S$ values on the resulting mass profile and the magnification profile.

By using the following lens equation (eq. \ref{lens}) one can lens
images back onto the source plane, and re-lens the source into the
lens plane:

\begin{equation}
\label{lens}
\vec\theta = \vec\beta + K_{q}\vec\alpha_T(\vec\theta),
\end{equation}

where $K_{q}$ is now scaled so it corresponds to
the observed locations of the images of each system.

\subsubsection[]{The Image Plane RMS}\label{sss:criteria1}

We relens the images of likely lensed galaxies in our search for counter images,
using all the pixel information as described in the beginning of \S \ref{model}. We measure
the RMS of the reproduced images with respect to their observed locations
to measure the fit:

\begin{equation} \label{RMS}
RMS_{images}^{2}=\sum_{i} (x_{i}^{'}-x_{i})^2 + (y_{i}^{'}-y_{i})^2 ~/ ~N_{images},
\end{equation}
where $x_{i}^{'}$ and $y_{i}^{'}$ are the locations given by the
model, and $x_{i}$ and $y_{i}$ are the real images location, and the
sum is over all $N_{images}$ images. Each system was minimised to its best reproduction.

The effect of different $q$ and $S$ on the image reproduction can be
generally described as follows: for a fixed smoothing degree $S$, a
shallow galaxy mass profile (lower $q$) generates too shallow a cluster
mass profile, resulting in a higher magnification that generally
predicts extra unobserved images (see Figures \ref{fig:mass_comp} to \ref{fig:mag_compS}). On the other hand, too high a value
of $q$ steepens the cluster mass profile, resulting in a lower
magnification and generally does not reproduce all the observed
images. For a fixed $q$, too low a smoothing degree $S$ produces
slightly too shallow a mass profile (meaning slightly higher
magnification), but more significantly the matter distribution is then
not detailed enough in its internal structure and therefore generally
underpredicts the number of observed images, or generates images with
incorrect orientation or twisted internal shape. A too high smoothing
degree $S$ produces a slightly too steep a mass profile and too
detailed a mass distribution, resulting in extra or missing images, or images with
twisted shapes.

\subsubsection[]{The Comparison to photo-$z$'s}\label{sss:criteria2}

We examine the growth of the model scaling factor $K_{q}$ versus the expected growth of $d_{ls}/d_{s}$ with
source redshift, based on the standard cosmological relation. The
difference between the predicted and measured redshifts is then characterised by:

\begin{equation} \label{RMSz}
RMS_{z}^{2}=\sum_{i} (z_{i}^{model}-z_{i}^{photo})^2 ~/ ~N_{systems}
\end{equation}

where the sum is over the number of systems used here,
$N_{systems}$. However, since the model-predicted redshifts are
manifested in the scaling factor $K_{q}$, this formula can be expressed in
terms of this factor:

\begin{equation} \label{RMSk}
RMS_{K_q}^{2}=\sum_{i} (K_{q,i}^{model}-K_{q,i}^{photo})^2 ~/ ~N_{systems}.
\end{equation}

The effect of varying the profile with $q$ and $S$, on the photo-$z$
predictions can be generally described as follows: for a fixed
smoothing degree $S$, a too shallow galaxy mass profile (lower $q$) results in a shallower mass profile (see Figure \ref{fig:mass_comp}) which
``compresses" the range of photo-$z$ compared to the cosmological
relation, so that we underestimate the redshift of systems which
exceed the reference z=1.675, for which $K_{q}>1$, and overestimate the
redshift of systems which lie below z=1.675, for which $K_{q}<1$. On the
other hand, choosing too high a value of $q$ corresponding to too
steep a profile has the opposite effect of ``expanding" the range of
predicted photo-$z$'s compared with the cosmological expectation.
Note that the exact choice of cosmology is not at all important, all
the predicted curves look alike, as unfortunately the variation of the
lensing distance ratio, $d_{ls}/d_s$, is relatively insensitive to
cosmology as shown in Figure \ref{fig:cosmo}.

\subsubsection[]{The Comparison of Relative Magnifications}\label{sss:criteria3}

The relative magnification of different images of a system also
provides a consistency check for the model. Since
gravitational lensing preserves surface brightness, both the total flux from each
image and the area of each image of a source are proportional to the
magnification at the image position. Therefore the ratio between the
fluxes (or the areas) of two images of a source should be the same as
the model magnification values at the image positions, and this can
be used to check the model.

In order to quantify the discrepancy of the relative magnification we
use the following procedure: we denote $F_{1}$ and $F_{2}$, as the
measured fluxes of two images of the same system, and $m_{1}$ and
$m_{2}$ as the magnification values at these image positions. By
definition, $F_{1}=F \cdot m_{1}$ and $F_{2}=F \cdot m_{2}$, where $F$
is the same original flux and it is therefore identical for both
images of the same system. The model-predicted magnitude difference of
these images ($\Delta mag = mag1-mag2$) is therefore given by:
\begin{equation}
\Delta mag = -2.5 (log m_{1}- log m_{2}).
\end{equation}
We use the latter equation to calculate the RMS of the relative magnification, by comparing the measured magnitude difference to the latter model-predicted value. This is obtained by:
\begin{equation} \label{RMSmag}
RMS_{mag}^{2}=\sum_{i} (\Delta mag^{meas.}-\Delta mag^{model})^2 ~/ ~N_{images},
\end{equation}
where the sum is over all images, and each magnitude difference is calculated with respect to the first image of the corresponding system of the image.

\begin{figure}
\centering{ \includegraphics[trim = 3cm 6cm 2cm 8cm, clip, width=8cm]{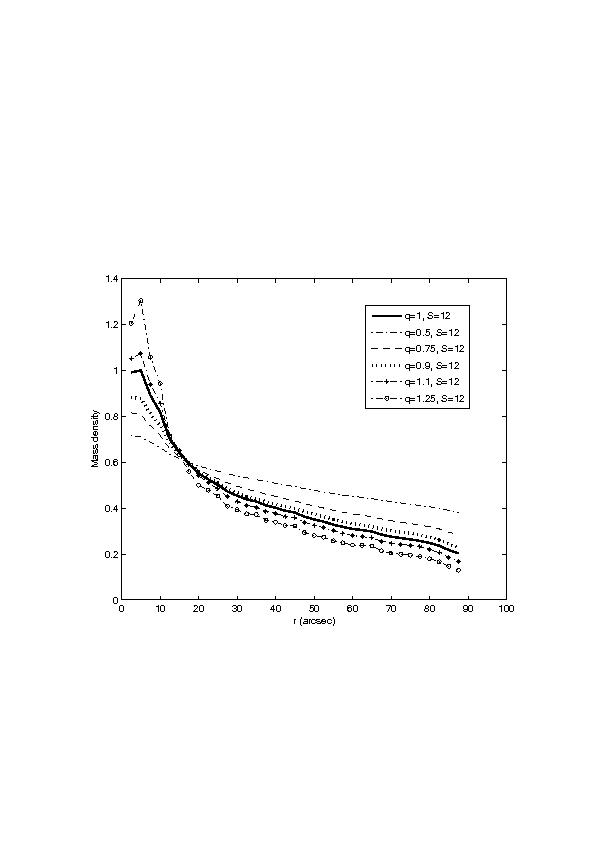}
  \caption{Comparison of different mass density profiles as a function of
  $q$, where the peak mass densities are normalised to the case $q=1$. The centre is slightly shifted to better
  see the limit each profile peaks to.}\label{fig:mass_comp}}
\end{figure}

\begin{figure}
\centering{ \includegraphics[trim = 3cm 8cm 2cm 8cm, clip, width=9cm]{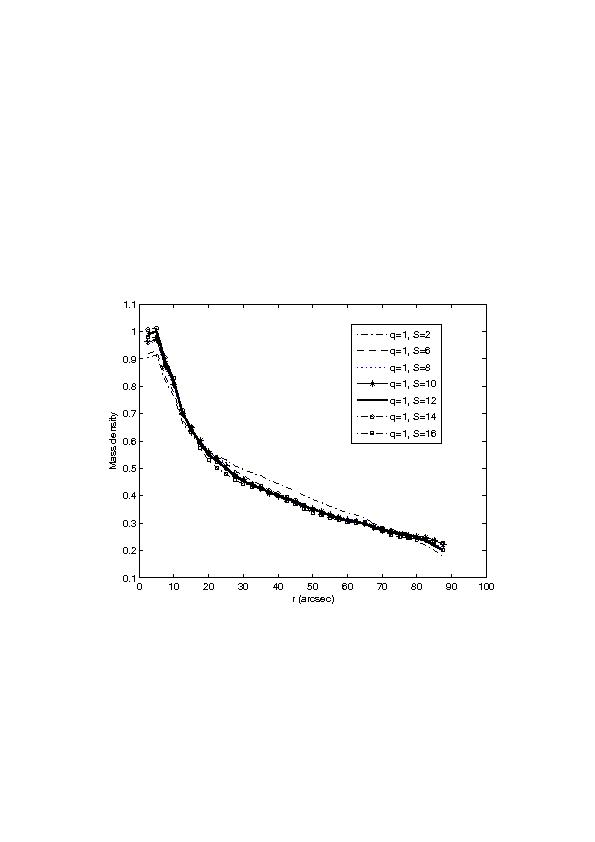}
  \caption{Comparison of different mass density profiles as a function of $S$ for fixed $q$.
   The mass density is normalised as before. The centre is slightly shifted to better
  see the limit each profile peaks to.}\label{fig:mass_compS}}
\end{figure}

\begin{figure}
\centering{ \includegraphics[trim = 3cm 8cm 2cm 8cm, clip, width=8cm]{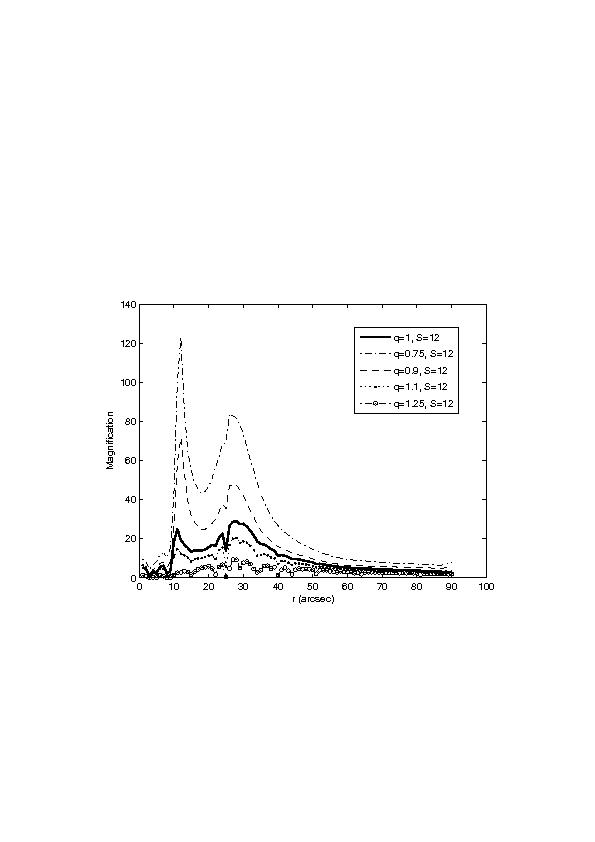}
  \caption{Comparison of different magnification profiles, as a function of $q$, for fixed $S$. Notice the radial and tangential
critical radii which are distinctive for low $q$ models.}\label{fig:mag_comp}}
\end{figure}

\begin{figure}
\centering{ \includegraphics[trim = 3cm 8cm 2cm 8cm, clip, width=8cm]{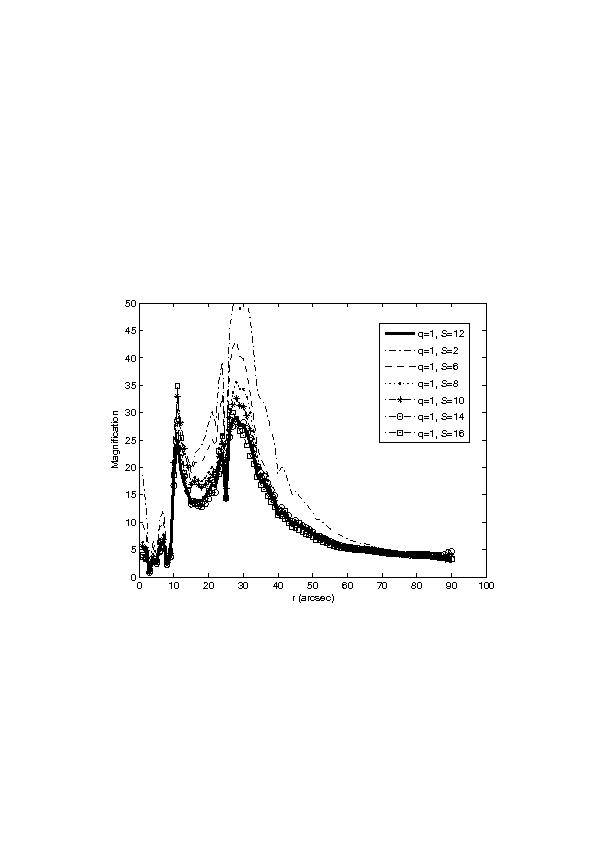}
  \caption{Comparison of different magnification profiles, as a function of $S$, for fixed $q$.}\label{fig:mag_compS}}
\end{figure}

As can be seen from Figures \ref{fig:mass_comp} to
\ref{fig:mag_compS}, the magnification is very sensitive to the mass
profile: relatively small changes in the mass steepness will result in
much higher changes in the magnification (and relative
magnifications). A higher $q$ and $S$ form a steeper profile for
which both the magnification and the relative magnifications are
higher, and vice versa. Due to the entanglement to the photo-$z$ comparison criterion,
the relative magnifications are used only as a consistency check of our
best model (see next subsection).


\subsection[]{Constraining the Mass Profile}\label{ss:realmodel}

We stressed in the preceding sections that only broad constraints on the
gradient of the mass profile are feasible using just the locations of the
multiply-lensed images alone, and that a tighter constraint
on the slope-determining parameters $q$ and $S$ comes from
incorporating the cosmological redshift-distance relation. This is done by applying
the lensing distances of each system based on the measured
photo-$z$'s.

We now wish to select the solutions for which the $d_{ls}/d_s$
model-estimates lie closest to the expected cosmological relation (see also \S \ref{sss:criteria2}). This will then allow us to define better the
gradient of the mass profile, as previously mentioned.
In so doing we concentrate in particular on the $z\simeq$4 system (system number 8),
which is our most distant multiple-image system and for which the
photometric redshift estimation is unambiguous and relatively
accurate due to the marked break in the SED (Figure \ref{fig:seds}), corresponding to
the Lyman limit and intervening Lyman forest absorption.

Finding the best fitting combinations of $q$ and $S$ is straightforward
since too shallow a slope leads to an underestimate of the redshift of
the $z\simeq$4 system, and too steep a slope overestimates it (as detailed in \S \ref{sss:criteria2}). This is because, as can be seen in Figures \ref{fig:mass_comp} and \ref{fig:mass_compS}, a higher $q$, and to a lesser extent a higher $S$, would result in a steeper overall mass profile, and vice versa. The best
fitting solution lies near $q$=1.3, $S$=10, with some
degeneracy between $q$ and $S$, which is not that physically important as
the parameters combine so that any solution near this position
generates a very similar mass profile (see Figure \ref{fig:Kappa}).

\begin{figure}
\centering{
  \includegraphics[trim = 2.2cm 1.5cm 2cm 13cm, clip, width=8cm]{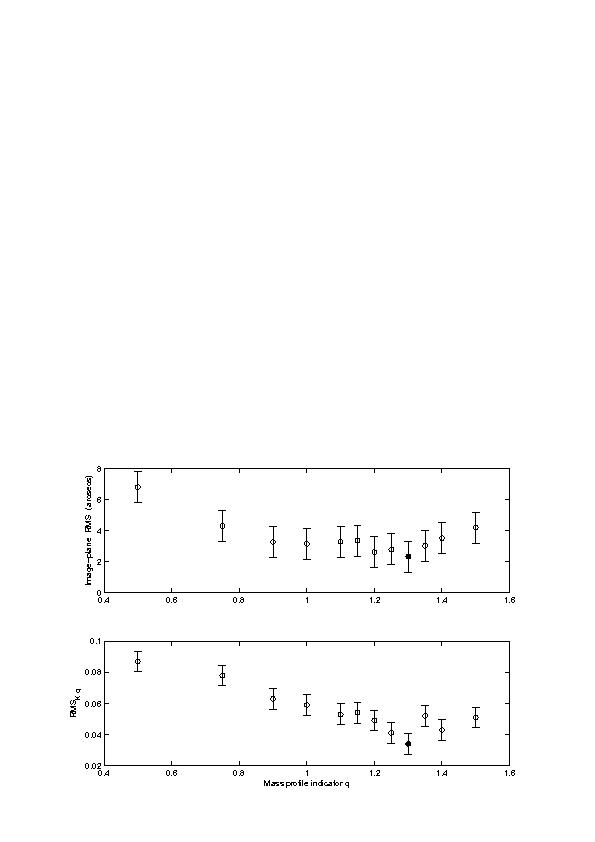}  \caption{Top: typical image-plane RMS as a function of $q$. Bottom: typical $RMS_{K_q}$ as a function of $q$. A minimum is seen around a value of $q\simeq1.3$ in both cases, but it is clear that the minimum is better defined by the use of photometric redshifts, in terms of $RMS_{K_q}$.}\label{fig:bestmodel}}
\end{figure}

\begin{figure}
\centering{ \includegraphics[trim = 3cm 8cm 2cm 8cm, clip, width=8cm]{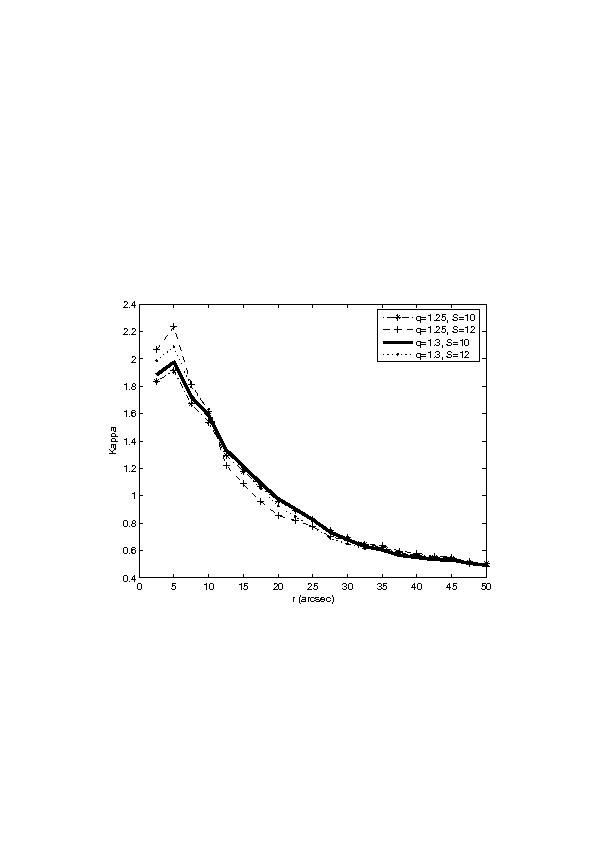} \caption{Mass profiles of the best-fitting model (solid curve). The dashed curves indicate the profiles of three other solutions which differ by $\sim1~\sigma$ from the best solution, in terms of both slope controlling parameters $q$ and $S$. The
  y-axis is the lens convergence, $\kappa$ (which is proportional to the mass density), where
  for a source of $z=1.675$ behind this cluster
  $\Sigma_{crit}=0.47~g/cm^2$. The x-axis is the radius from the cD
  galaxy in arcseconds, where each arcsecond equals 5.3
  kpc/$h_{70}$.}\label{fig:Kappa}}
\end{figure}

We then examine how well the cosmological relation is reproduced accounting also for the other systems with reliable
photo-$z$ measurements (9 in total), as shown in Figure
\ref{fig:bestmodel} (bottom). Note that the best fitting model as obtained from system 8 above is also in best agreement with the redshift estimates of
all the 9 systems together. Clearly the redshift estimates for these systems verify very well that the predicted deflection ($K_q$) of the
best fitting model at the estimated redshifts of each of these 9 systems
lies precisely along the expected cosmological relation, as seen in Figure \ref{fig:cosmo}, with a mean deviation of only $RMS_{K_q} \sim0.03$
for the best model, considerably strengthening the
plausibility of our approach to modelling in general. Note that if we rely only on the image locations, Figure \ref{fig:bestmodel} (top), although we obtain a similar best-fitting value of $q$, the minimum is not as clear.

In Table \ref{systems} we also include the relative magnifications
comparison of the best model as a consistency check. Here we also see
good consistency with the best fitting model ($RMS_{mag}\sim0.5~mag$), with the data lying very
closely along the equality slope (Figure \ref{fig:relmag}), with the scatter
increasing symmetrically with increasing magnitude. The most discrepant
point (lower right) is one of the images of system 7 for which the
photometry is difficult due to its anomalously low surface brightness and for
which the predicted redshift, given the cosmological relation, is not
matched by our photo-$z$ estimated redshift.

We now show in the next section a
comparison between the predicted images and their model-generated
images (for the best-fitting model), in order to demonstrate the plausibility of our solution both
in terms of the appearances of the images and their photometric
redshifts.

\section[]{Results and discussion}\label{Results}
In this section we present the new multiple-image systems found by the
model, the derived mass distribution and profile, and the resulting magnification. We also make consistency checks of our
model using independent information including photometric redshifts,
relative fluxes, WL data, and a comparison to an NFW profile.

\subsection[]{Multiply-Lensed Systems and Candidates}
Here we discuss the various multiply-lensed systems. Each system
identified with the help of the model is presented and accompanied by
modelled images of the same system. Each image is relensed using all the pixel information contained in
one of the counter images, so that each modelled image is the relensed counter-image of another image of the same system. This detailed modelling of the images is shown together with the photometric redshift
probability function of each image, so that the precision and
uniqueness of the photometric redshift can be assessed.

All systems are summarised in Table \ref{systems}, and are marked on a
colour image along with the critical curves (see Figure
\ref{fig:MagMap}). The model reproduces quite accurately nearly all lensed galaxy images, with respect to
their locations ($RMS\sim2.5$\arcsec) and redshift ($RMS_{K_q}\sim0.03$), as well as their shape and relative magnification ($RMS_{mag}\sim0.5~mag$). In the few cases where there is
ambiguity in one or more of the counter images we present the alternative
possibility together with the favoured identification. It should be
noted however that this degeneracy does not alter the mass
distribution as these alternative counter images lie very close to
each other.

\begin{figure}
\centering{
  \includegraphics[trim = 3cm 7cm 1cm 5cm, clip, width=8cm]{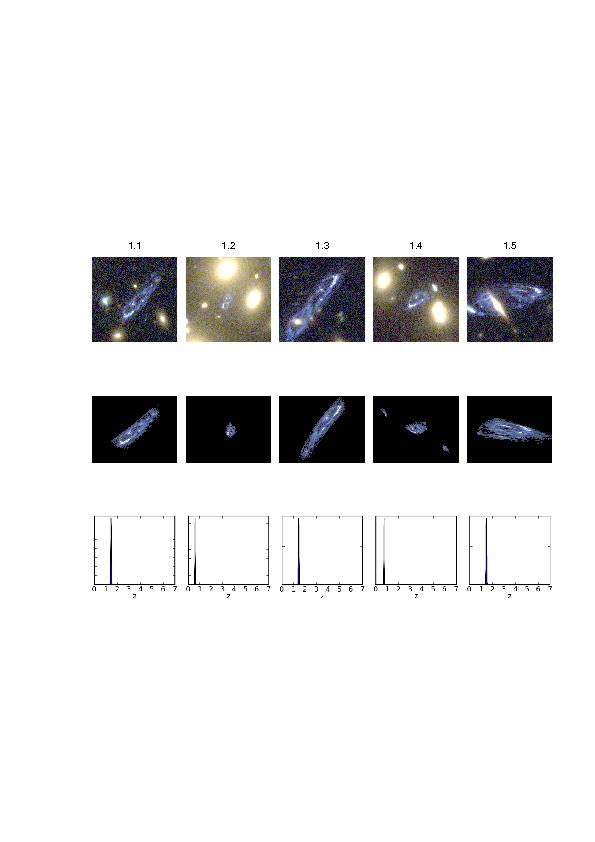}
\caption{System number 1: Five images of the well known ``Theta"-arc are
displayed in the top row (1.1 to 1.5), including the central
radial arc, 1.2, with the best-fitting model reproduction shown
below. All model images are based on our delensing-relensing
technique, allowing for the most detailed comparison possible given
the resolution of the data. The reproduction of images 1.3 and 1.5 is
somewhat complicated by the presence of nearby cluster member
galaxies, whose mass profiles
will be interesting to explore further. The photometrically determined redshift probability distribution
based on our optical-IR measured photometry is shown below. Note that
images 1.2 and 1.4 are significantly affected by nearby light from
cluster member galaxies (see Figure \ref{fig:seds}), badly affecting their
photometric redshift estimation, whereas image 1.1, 1.3 and 1.5 predict
a photometric redshift of, $z=1.45\pm0.25$, (Table \ref{systems}) in agreement
with the measured redshift of $z=1.675$ (Broadhurst et al.
2000).}\label{fig:1_z}}
\end{figure}

${System 1}$: This is the historically well-known system,
consisting of five images, first noted by Koo (1988). Three of these
images were resolved into a close triplet of arcs (Kassiola, Kovner,
\& Blandford, 1992; images 1.3, 1.4, and 1.5 in Figure \ref{fig:1_z}
here), where the two additional images (1.1 and 1.2) were later also
classified as members of this system (e.g., Smail et al. 1996, Colley,
Tyson \& Turner 1996). This system is the only multiply-lensed system
in this cluster which has a spectroscopic redshift ($z=1.675$,
Broadhurst et al. 2000). We therefore use this system for the internal
minimisation of the model, which sets the scaling factor $K_{q}$ to 1 for
this system. In Figure \ref{fig:1_z} we show the five images of this
system in the ACS colour image, their reproduction by the model, and
the photo-$z$ distribution of each image. The images are accurately
reproduced by the best fitting model (Figure \ref{fig:1_z}). Note that
next to image 1.4 two small spots are reproduced but if they are
realistic and not a lensing noise effect, they are hidden behind the
bright galaxies seen in the same figure. The photo-$z$'s of the bright
images (1.1, 1.3 and 1.5) agree well with each other
($z\simeq1.45\pm0.25$, see Table \ref{systems}), and with the
spectroscopic redshift ($z=1.675$). The two fainter images (1.2 and
1.4), both lie next to bright cluster galaxies, influencing their
photometry and therefore their photo-$z$ is unreliable (Table
\ref{systems}).

\begin{figure}
\centering{
\includegraphics[trim = 4cm 9cm 2cm 6cm, clip, width=8cm]{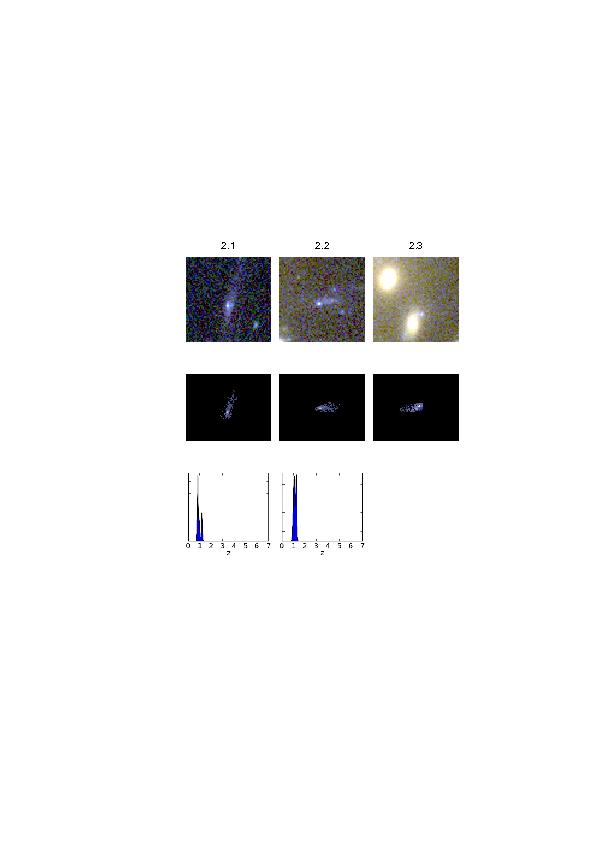}
\caption{System number 2: The three images of this system are shown in the top row, where below them we add the model-reproduced images. The third row is the photo-$z$ distributions of the images which agree with our $z\sim1.2$ prediction. Note that image 2.3 lacks photometric redshift as it is hidden in a bright cluster galaxy. Images 2.1 and 2.2 were first identified as a multiply-lensed system by Broadhurst et al. (2000), where image 2.3 was predicted recently by Liesenborgs et al. (2008b), and was found here (see also Figure \ref{fig:2_sub}). Note that images 2.2 and 2.3 lie on two sides of the radial critical curve and show opposite parity.}\label{fig:2_z}}
\end{figure}

\begin{figure}
\centering{
  \includegraphics[trim = 4cm 2cm 4cm 14cm, clip, width=8cm]{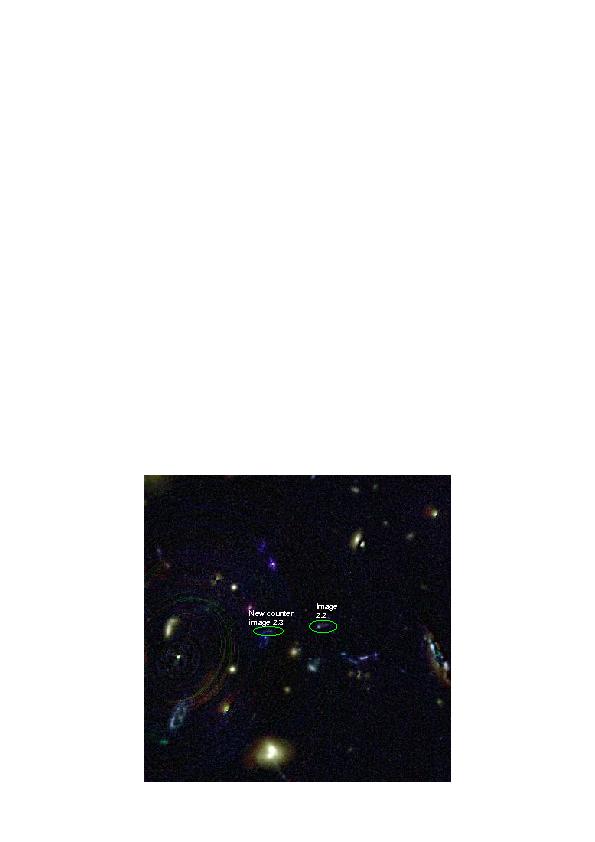}
\caption{Newly-revealed radial image of system number 2, along with its
counter radial-image. The cluster galaxies were subtracted from the image,
revealing image 2.3.}\label{fig:2_sub}}
\end{figure}

${System 2}$: We identify three images for this system (Figure
\ref{fig:2_z}), two of which were first proposed by Broadhurst et
al. (2000) based on shallow WFPC-2 images. These images were later
observed to have very similar internal colours and structure
in deep ACS images and were used by Jee et al. (2007). Our model confirms
these images to be members of the same system and predicts a third
radial image, 2.3, lying just inside the radial critical curve, revealed for the first time in a galaxy-subtracted image made by us (Figure \ref{fig:2_sub}). This additional radial image was predicted also in recent work by Liesenborgs et al. 2008b. The SEDs of images 2.1 and 2.2 are
very similar (Figure \ref{fig:seds}). Images 2.2 and 2.3 are radially
directed, and have opposite parity, indicating that they lie on
opposite sides of the radial critical curve in agreement with our model
prediction.
Our best fitting model predicts a redshift of $z=1.22\pm0.1$ for this
system in
good agreement with the estimated
value of $z=1.24$ by our photo-$z$ analysis, combining the estimates of
images 2.1. and 2.2.

\begin{figure}
\centering{
  \includegraphics[trim = 3cm 7cm 1cm 5cm, clip, width=8cm]{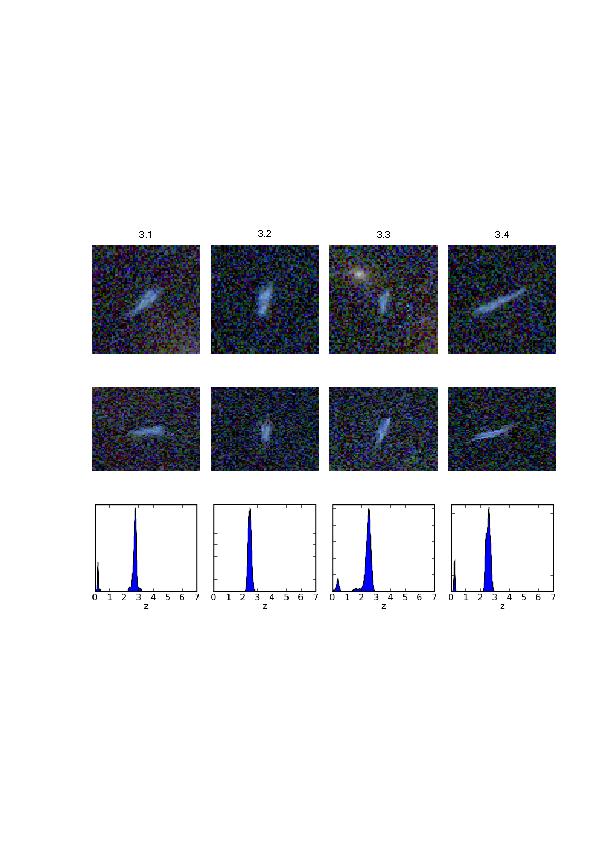}
\caption{System number 3: The four images of this system are shown in the top row, where below them we add the model-reproduced images on similar artificial background. The third row is the photo-$z$ distributions of the images which agree with our $z\sim2.6$ prediction.}\label{fig:3_z}}
\end{figure}

${System 3}$: This system comprises four blue images with a
similar lumpy internal substructure. The predicted redshift from our
lensing
model is $2.55^{+0.45}_{-0.20}$ where photo-$z$'s of these images
range from $z=2.48$ to $z=2.76$ (Table \ref{systems}), in good
agreement with the model. One image (3.1 in Figure \ref{fig:3_z})
appears in the south-eastern side of the cluster, opposite to the
three other images on the north-western side. These images have very
similar SEDs as can be seen in Figure \ref{fig:seds}. It should be
noted that images 3.1 and 3.4 are close to images 5.2 and 5.1,
respectively, and show similar colours (to the eye). This might imply a
certain degeneracy between these systems. However, an inspection of
these systems SEDs (Figure \ref{fig:seds}) and photo-$z$ distribution,
supports the model-prediction that these systems are multiply-lensed
in the form presented here. In addition, the system 5 SEDs were compared to the mean system 3 SED.
The former are steeper and, based solely on the photometric uncertainties,
disagree with the latter with $\chi^2 = 3.4, 3.6$ ($\sim 93\%$ confidence). Contamination from nearby neighbors may also contribute to SED disagreement in general,
though these galaxy images appear fairly well isolated.

\begin{figure}
\centering{
  \includegraphics[trim = 4cm 9cm 2cm 6cm, clip, width=8cm]{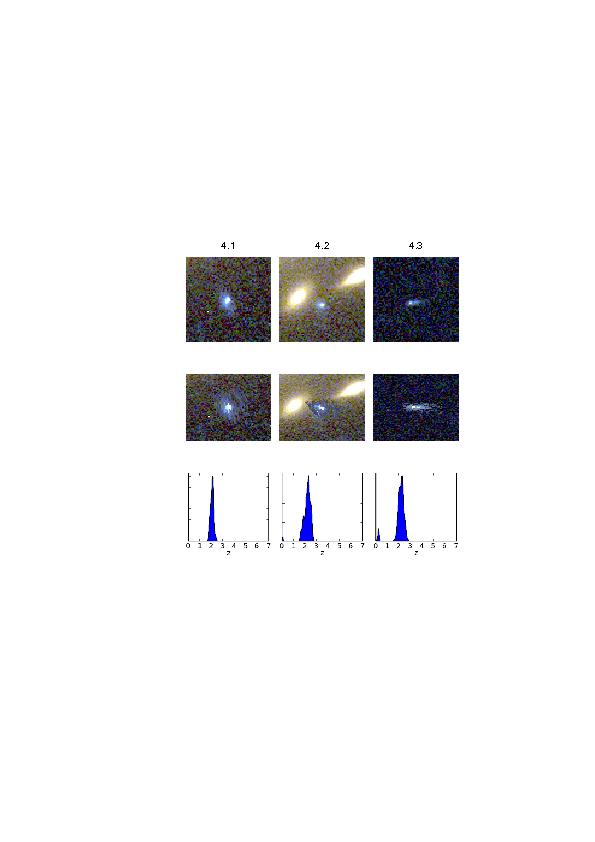}
\caption{System number 4: The three images of this system are shown in the top row, where below them we add the model-reproduced images. The third row is the photo-$z$ distributions of the images which agree with our $z\sim2$ prediction. Images 4.1 and 4.2 lie on the two sides of the radial critical curve, and are locally lensed by cluster members which further magnify them.}\label{fig:4_z}}
\end{figure}

${System 4}$: Images 4.1 and 4.2 are very close to each other and
relatively close to the cluster centre, and lie on the opposite sides
of the radial critical curve next to cluster galaxies, and therefore
are highly magnified. The third image, 4.3, is on the other side of
the cluster, and it is less magnified, as can be seen in Figure
\ref{fig:4_z}. The internal structure is very well reproduced by our
model, in particular the location of the white knot relative to the
diffuse blue emission. The lensing model predicts a redshift of
$1.96\pm0.2$, similar to the photo-$z$ estimate of $z\sim2.2$. Note that for image 4.1 there is no IR data since
it lies outside the region covered by NIC3. The SEDs of
images 4.1, 4.2, and 4.3 (Figure \ref{fig:seds}) have a similar shape
in the optical, but image 4.2 is slightly brighter in the red, due to
contamination by nearby cluster members.

\begin{figure}
\centering{
  \includegraphics[trim = 3cm 7cm 3cm 6cm, clip, width=8cm]{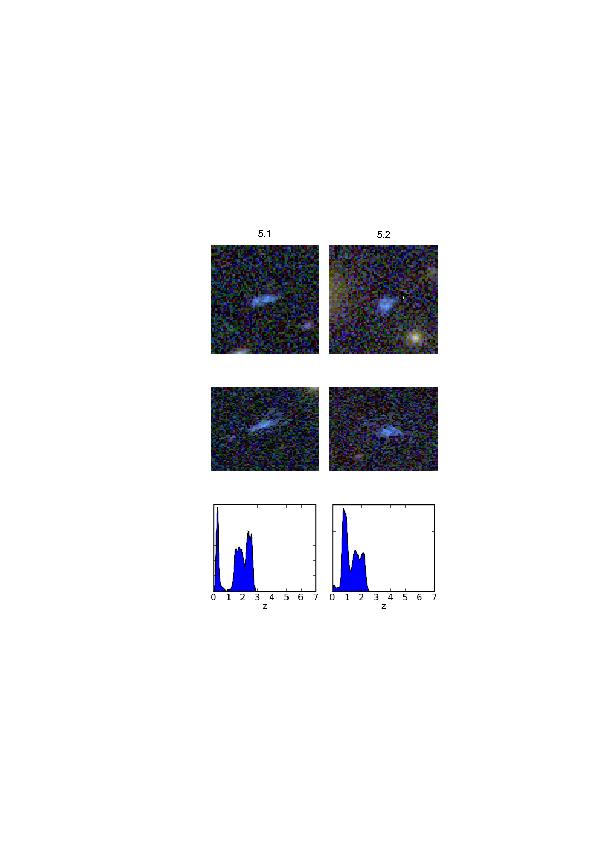}
\caption{System number 5: The two images of this system are shown in the top row, where below them we add the model-reproduced images on similar artificial background. The third row is the photo-$z$ distributions of the images which agree with our $z\sim2$ prediction.}\label{fig:5_z}}
\end{figure}

${System 5}$: Two blue images, seen in two different sides of the
cluster. Both images have similar structure as can be seen in Figure
\ref{fig:5_z}, and the model-predicted counter images generated by
relensing each of these images for comparison with the other, look
very much like the observed images. Also, both images have similar SEDs
(Figure \ref{fig:seds}). Image 5.1 lies right on the
boundary of the IR coverage and so we restrict the photo-$z$ estimate
only to the optical data in both cases. This produces a photo-$z$
of $z=2.02$ for both images, which accurately fits the
lensing model which predicts $z=2.02^{+0.10}_{-0.10}$. Note that this redshift agrees also with the IR-included photo-$z$ as it is bimodal (see Figure \ref{fig:5_z}).

\begin{figure}
\centering{
  \includegraphics[trim = 3cm 7cm 3cm 6cm, clip, width=8cm]{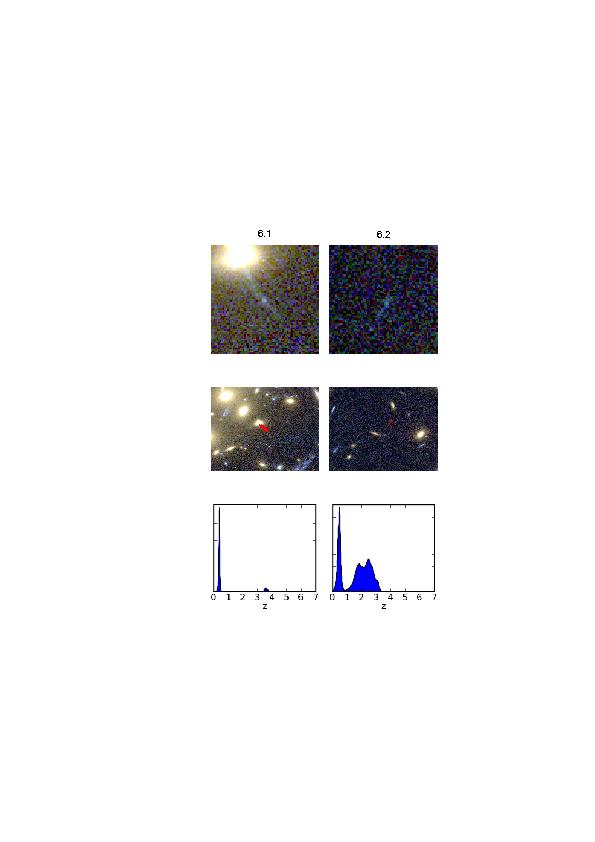}
\caption{System number 6: The two images of this system are shown in the top row, where below them we add the model-reproduced images. The third row is the photo-$z$ distributions of the images. Note that due to the vicinity of image 6.1 to a bright cluster galaxy, its photo-$z$ estimate is unreliable. The photo-$z$ distribution of image 6.2 agrees with our model estimate of $z\sim1.9$. Also, due to the vicinity of image 6.1 to a cluster member, the colour delensing-relensing is seriously contaminated. We therefore lens each image of this system to reproduce its counter image, painted in red to emphasise the accuracy of the model with respect to images location. This is seen in the second row, where we also zoom out in order to better show the relative location of the reproduced images in the cluster.}\label{fig:6_z}}
\end{figure}

${System 6}$: Image 6.1 is a radial arc next to a bright central
cluster
member, whereas its companion is revealed on the
other side of the cluster. Due to the proximity to the cluster member,
the SED of image 6.1 is badly affected. The photo-$z$ distribution of
image 6.2 is bimodal, with the higher redshift peak at $z\simeq2$
in agreement with our lensing model which predicts
$z=1.93^{+0.15}_{-0.12}$ for this system, whereas the lower
redshift photo-$z$ peak at $z=0.46$ is too low to produce a counter
image and therefore excluded. Note that in Jee et al. (2007), the radial arc (image 6.1 in Figure
\ref{fig:6_z}) was proposed to match with a counter image on the same
side of the lens as image 6.1, which is physically impossible in the
context of our model. It should also be noted that there are two similar looking arcs seen in the vicinity of image 6.2, yet these are excluded via SEDs and photo-$z$ comparison.

\begin{figure}
\centering{
  \includegraphics[trim = 3cm 7cm 3cm 6cm, clip, width=8cm]{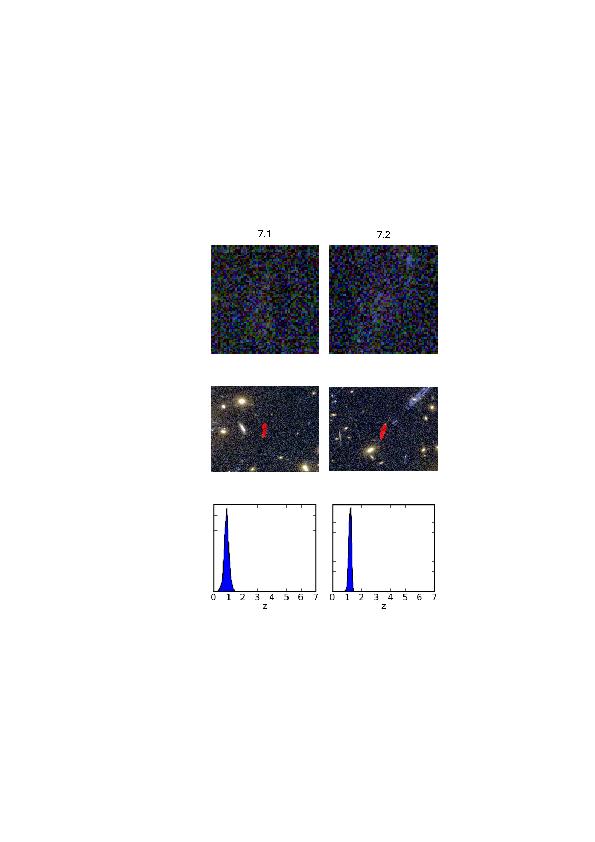}
\caption{System number 7: The two images of this system are shown in the top row, where below them we add the model-reproduced images. The third row is the photo-$z$ distributions of the images. Note that due to the low surface brightness of these images, their photo-$z$ estimate is unreliable. Our model reproduces these images with a source redshift of $z\sim2.2$. Being too faint to notice here, we lens each image of this system to reproduce its counter image, painted in red to emphasise the accuracy of the model with respect to images location and general shape. This is seen in the second row, where we also zoom out in order to better show the relative location of the reproduced images in the cluster.}\label{fig:7_z}}
\end{figure}

${System 7}$: Two large, low-surface-brightness and slightly reddish
images, lying on opposite sides of the cluster. Though somewhat too faint to
be easily noticed in Figure \ref{fig:7_z}, a close inspection by eye
in a high resolution colour image shows the colour and shape
resemblance. The lensing model reproduced the locations and shape of
these images well, at a best fitting redshift of
$z=2.15^{+0.28}_{-0.26}$. However, the photometric redshift is not in
close agreement with this prediction low, $z\sim1$, see Table
\ref{systems}. The SEDs of these images are relatively similar but
with relatively large photometric errors (due to the images being
faint). We do not trust the photo-$z$ prediction in that case and
conclude that it is plausible that the rather red SED of this system
may be significantly affected by extinction, and that for such objects we
lack
suitable templates for a reliable photo-$z$ determination.

${System 8}$: Two green images, the first (8.1) is a tangential
arc $\sim50"$ from the cluster centre. The second (8.2) falls on the
other side of the cluster, closer to the centre. Photometric redshifts
predict $z=4.09$ and $z=4.16$ for these images respectively, in
agreement with our lensing model which predicts a similar redshift of
$z=4.03\pm 0.5$. Also, the SEDs of these images look very similar (Figure
\ref{fig:seds}) with the characteristic break expected of a distant
drop-out galaxy. Note that one image (8.1) lacks NIC3 coverage.
\begin{figure}
\centering{
  \includegraphics[trim = 3cm 7cm 3cm 6cm, clip, width=8cm]{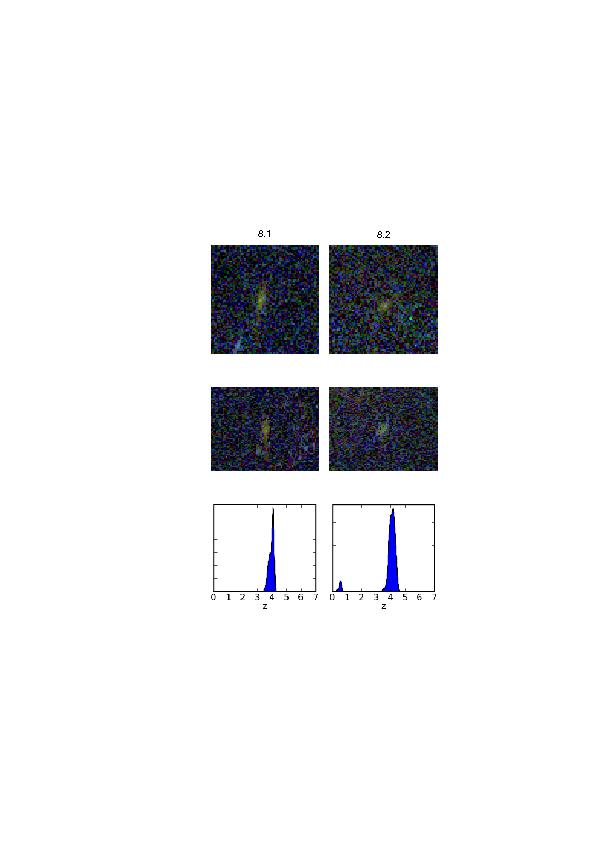}
\caption{System number 8: The two images of this system are shown in the top row, where below them we add the model-reproduced images on similar artificial background. The third row is the photo-$z$ distributions of the images which agree with our $z\sim4$ model prediction. Note that this system has the highest source redshift among the images found here as its images show the characteristic break expected from a distant drop-out galaxy, and it was used to strongly constrain the mass profile, as detailed in \S \ref{ss:realmodel}.}\label{fig:8_z}}
\end{figure}

\begin{figure}
\centering{
  \includegraphics[trim = 4cm 9cm 2cm 6cm, clip, width=8cm]{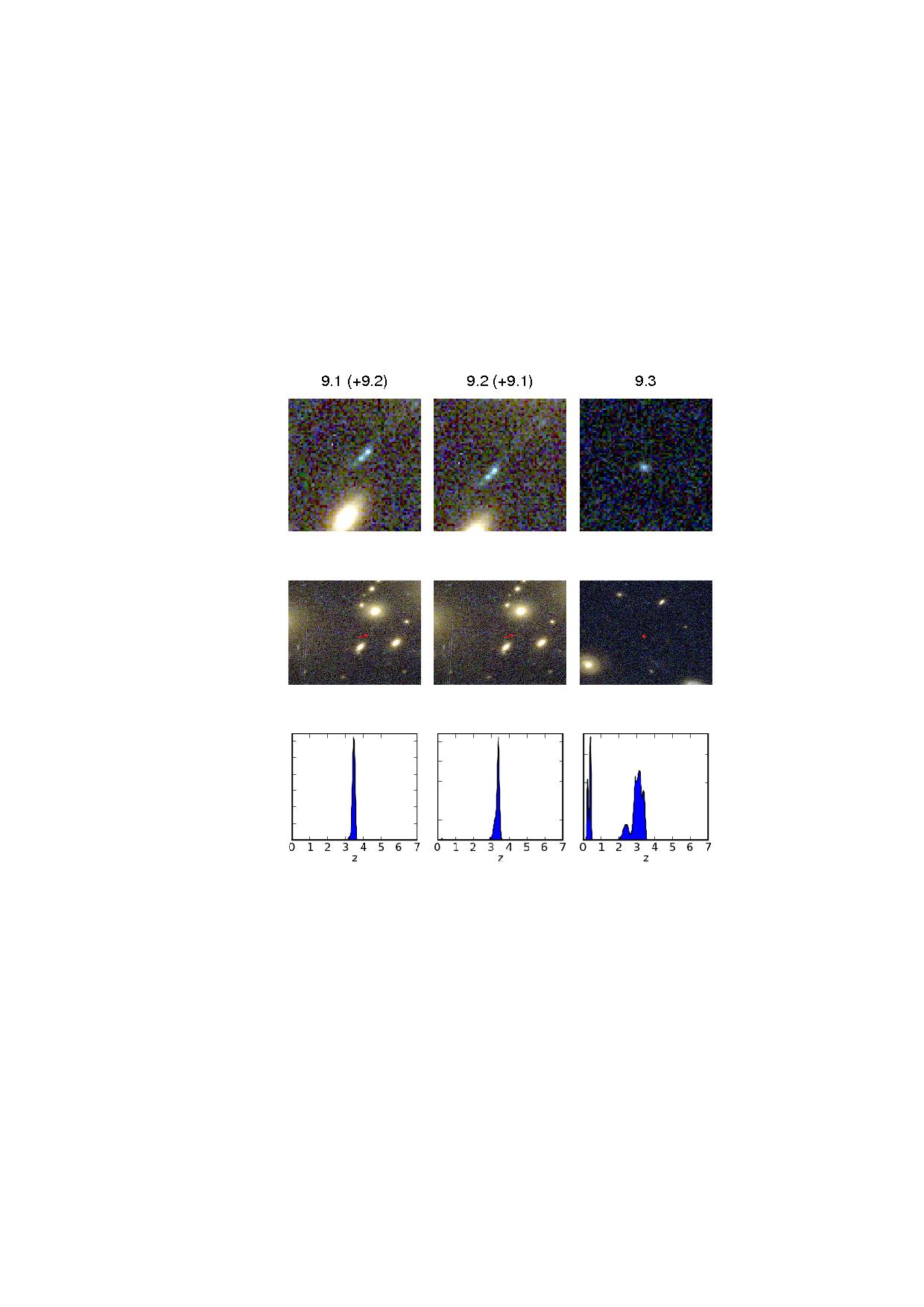}
\caption{System number 9: The three images of this system are shown in the top row, where below them we add the model-reproduced images. The third row is the photo-$z$ distributions of the images. Note that the photo-$z$ of images 9.1 and 9.2 is badly affected by galaxy light, and we rely on the photo-$z$ distribution of image 9.3, which agrees with our $z\sim2$ model prediction. Images 9.1 and 9.2 are, as shown by our model, two different images, straddling the radial critical
curve. Also, due to the vicinity of images 9.1 and 9.2 to a cluster member, the colour delensing-relensing procedure is seriously contaminated. We therefore lens each image of this system to reproduce its counter image, painted in red to emphasise the accuracy of the model with respect to the images location. This is seen in the second row, where we also zoom out in order to better show the relative location of the reproduced images in the cluster. Note that the orientation of the combined image 9.1+9.2 is slightly different than the original as it is strongly locally affected by the nearby cluster galaxy.}\label{fig:9_z}}
\end{figure}

${System 9}$: Images 9.1 and 9.2 (Figure \ref{fig:9_z}) are
probably two different images, straddling the radial critical
curve. Their counter image (9.3) is a small bright spot, which has a
similar SED (Figure \ref{fig:seds}). This configuration is given for a
source redshift of $z\sim2$, according to the lensing model. Note that
this does not agree with the photo-$z$ estimation for images 9.1 and
9.2 ($z\sim3.4$), which is significantly affected by the nearby
bright cluster member. The photo-$z$ distribution of image 9.3 is
bimodal, with the higher redshift option lying in the range
$z\sim2-3.5$, in agreement with our lensing model. Note that in the
vicinity of image 9.3 there is another small arc with a bright spot
(RA=00:26:37.85, DEC=+17:09:58.54), which has similar SED and might be another option, though it is less supported by our model.

\begin{figure}
\centering{
  \includegraphics[trim = 4cm 9cm 2cm 6cm, clip, width=8cm]{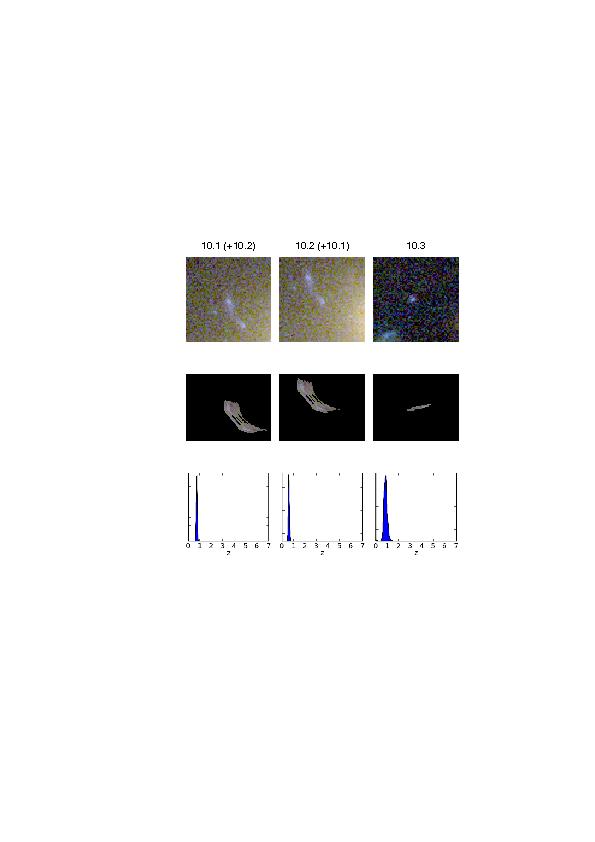}
\caption{System number 10: The three images of this system are shown in the top row, where below them we add the model-reproduced images. The third row is the photo-$z$ distributions of the images. Note that the photo-$z$ of images 10.1 and 10.2 is badly affected by galaxy light, and we rely on the photo-$z$ distribution of image 10.3, which agrees well with our $z\sim1$ model prediction. Images 10.1 and 10.2 are radial images straddling the radial critical curve. Also, due to the vicinity of these images to cluster members, the colour delensing-relensing procedure is relatively noisy, but it is sufficient to show the relevant internal structure. Note that the relensed images are somewhat bigger, as some of the noise gets lensed as well.}\label{fig:10_z}}
\end{figure}

${System 10}$: Images 10.1 and 10.2 (Figure \ref{fig:10_z}) are
two purple-like radial arcs in the cluster centre, probably counter
images of one another, lying across the radial critical curve. A third
counter image, 10.3, is found on the other side of the cluster. The SEDs
and photo-$z$ estimation of images 10.1 and 10.2 are affected by
the halo of bright cluster members. The photo-$z$ of image 10.3 is
0.85$^{+0.31}_{-0.26}$, in good agreement with our lensing model
prediction of $z=0.96^{+0.23}_{-0.20}$.

${System 11}$: This system imitates the location of the main
5-image system, with a slightly larger radius (and therefore redshift)
comprising four blue images with internal bright spots, next to images
1.1, 1.3, 1.4 and 1.5, respectively, and well reproduced by our
model. A very small image is predicted in the cluster centre near 1.2,
but was not found, probably due to the bright member light and
uncertain subtraction at the predicted location. All images show very similar SEDs. The model predicts a redshift of z=$2.50\pm0.5$
for this system, where photo-$z$ analysis shows an agreeing redshift of
$z\sim2.8$.

\begin{figure}
\centering{
  \includegraphics[trim = 3cm 7cm 1cm 5cm, clip, width=8cm]{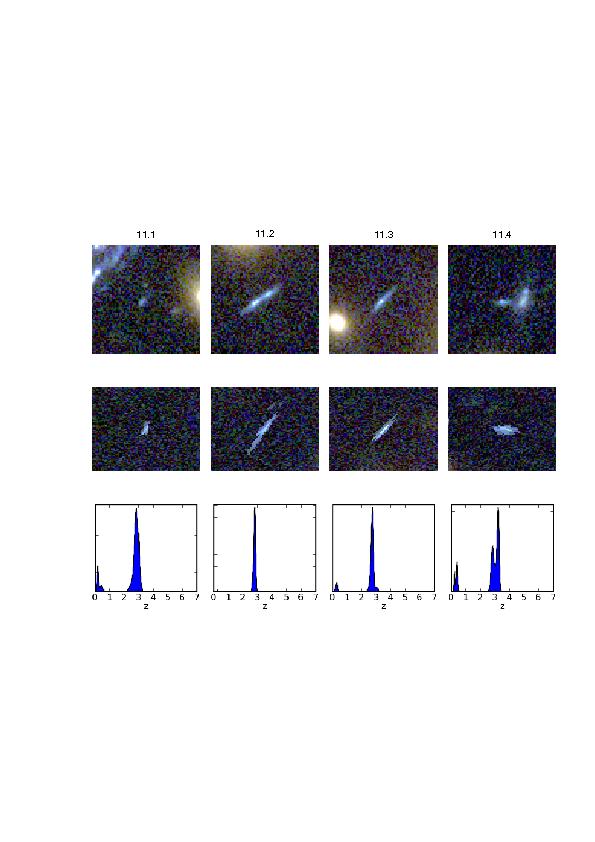}
\caption{System number 11: The four images of this system are shown in the top row, where below them we add the model-reproduced images on similar artificial background. The third row is the photo-$z$ distributions of the images, which agree well with the our model $z\sim2.5$ prediction. This system imitates the location of the main
5-image system, with a larger radius (and therefore redshift)
comprising four blue images with internal bright spots, next to images
1.1, 1.3, 1.4 and 1.5, respectively, and well reproduced by our
model. A very small image is predicted in the cluster centre near 1.2,
but was not found, probably due to the bright member light and
uncertain subtraction at the predicted location.}\label{fig:11_z}}
\end{figure}

\begin{table*}
\caption{Known and new multiple-images systems discovered by the model. For
more detailed information on each system and other optional members
please see the corresponding subsection. The columns are: arc ID; RA
and DEC in J2000.0; the $i'_{775}$ magnitude in the AB system;
relative magnification - these values represent the lensing-model
predicted magnitude difference, between the first image of each system
and each of the following images, based on the relative magnifications
(see \S \ref{sss:criteria3}); the lensing-model predicted
redshifts; the photometric redshifts, as obtained from the highest
peak photo-$z$ distribution and errors. Note that some images were not
covered by the NIC3 imaging and therefore their photo-$z$ relies on
ACS photometry alone. Note also that the errors are very large in
cases where the photo-$z$ distribution is bimodal (see images 6.1 and
9.3 in Figures \ref{fig:6_z} and \ref{fig:9_z}, respectively). In such
cases the values which agree with the SL model can be different than
specified in this Table, as they arise from other peaks in the
distribution, as detailed in the corresponding paragraphs; the
$\chi^2$ of the photometric redshift fitting procedure.}

\label{systems}
\begin{center}
\begin{tabular}{|c|c|c|c|c|c|c|c|}
\hline\hline
ARC & RA & DEC & $i'_{775}$ & Relative&&&\\ ID& (J2000.0)&(J2000.0)&
(AB mag)& magnification&$z_{model}$& $z_{phot}$& $\chi^2$\\
\hline
1.1 &00 26 34.43 &17 09 55.4 &21.333$\pm$0.008& ---&--- &1.44 $\pm$ 0.24 &0.29\\
1.2 &00 26 35.45 &17 09 43.1 &21.152$\pm$0.009& -0.229&" &0.60 $\pm$ 0.16 &0.26\\
1.3 &00 26 37.19 &17 09 15.8 &20.641$\pm$0.006 &-0.207& " & 1.45 $\pm$ 0.24 &0.41\\
1.4 &00 26 37.72 &17 09 25.7 &21.264$\pm$0.007& 0.290&" &0.71 $\pm$ 0.17 &0.68\\
1.5 &00 26 37.99 &17 09 41.2 &20.988$\pm$0.007& -0.458&" &1.46 $\pm$ 0.24 &0.39\\
2.1 &00 26 35.57 &17 09 08.8 &24.572$\pm$0.034& --- &1.22$\pm$0.1& 0.85$^{+0.41}_{-0.18}$ &0.29\\
2.2 &00 26 35.79 &17 09 52.7 &24.734$\pm$0.041& -0.467 &" &1.24$^{+0.22}_{-0.31}$ &0.17\\
2.3 &00 26 35.82 &17 09 49.0 &---& --- &" &--- &---\\
3.1 &00 26 36.86 &17 09 24.3 &25.142$\pm$0.039&--- &2.55$^{+0.45}_{-0.2}$& 2.76$^{+0.37}_{-2.59}$ &2.60\\
3.2 &00 26 35.66 &17 10 18.4 &25.176$\pm$0.039& -0.201&" &2.48 $\pm$ 0.34 &0.15\\
3.3 &00 26 34.74 &17 10 12.6 &25.953$\pm$0.071& 0.200&" &2.51$^{+0.34}_{-2.19}$ &0.26\\
3.4 &00 26 32.92 &17 09 46.6 &25.070$\pm$0.040& -0.005&" &2.58$^{+0.35}_{-2.38}$ &0.53\\
4.1 &00 26 34.59 &17 09 42.1 &24.454$\pm$0.028& ---&1.96 $\pm$ 0.20 &2.13$^{+0.31}_{-0.33}$ &0.15\\
4.2 &00 26 34.75 &17 09 41.5 &24.608$\pm$0.036&0.242 &" &2.30$^{+0.34}_{-0.68}$ &0.20\\
4.3 &00 26 38.73 &17 09 38.6 &26.088$\pm$0.060&0.403 &" &2.28$^{+0.36}_{-2.07}$ &0.74\\
5.1 &00 26 32.86 &17 09 39.3 &26.168$\pm$0.083& ---&2.02 $\pm$ 0.20 &0.25$^{+2.44}_{-0.12}$ &1.23\\
5.2 &00 26 36.76 &17 09 31.9 &26.083$\pm$0.068&0.130&" &1.58$^{+0.65}_{-1.52}$ &1.40\\
6.1 &00 26 35.13 &17 09 49.4 &23.494$\pm$0.038& --- &1.93$^{+0.15}_{-0.12}$ &0.38$^{+3.23}_{-0.13}$ &0.17\\
6.2 &00 26 37.78 &17 09 04.7 &27.444$\pm$0.173&-0.159 &" &0.46$^{+2.59}_{-0.18}$ &1.44\\
7.1 &00 26 35.58 &17 10 05.9 &25.587$\pm$0.158& ---&2.15$^{+0.28}_{-0.26}$& 0.90$^{+0.33}_{-0.38}$ &1.05\\
7.2 &00 26 36.22 &17 09 03.0 &25.137$\pm$0.116&-0.019&" &1.23$^{+0.22}_{-0.23}$ &0.71\\
8.1 &00 26 36.38 &17 08 55.9 &25.755$\pm$0.052&---  &4.03$\pm$0.5 &4.09 $\pm$ 0.50&0.29\\
8.2 &00 26 35.01 &17 10 02.8 &26.592$\pm$0.082& 0.182&" &4.16$^{+0.51}_{-3.62}$ &0.43\\
9.1 &00 26 34.66 &17 09 28.9 &26.296$\pm$0.068& ---&1.96$^{+0.5}_{-0.1}$& 3.46 $\pm$ 0.44 &0.83\\
9.2 &00 26 34.67 &17 09 29.2 &26.211$\pm$0.064& -0.007&" &3.40 $\pm$ 0.43 &0.81\\
9.3 &00 26 37.53 &17 10 06.8 &26.568$\pm$0.076& 0.033&" &0.38$^{+3.05}_{-0.17}$ &0.81\\
10.1 &00 26 36.17 &17 09 42.3 &24.703$\pm$0.034& ---& 0.96$^{+0.23}_{-0.20}$& 0.75 $\pm$ 0.17 &1.35\\
10.2 &00 26 36.11 &17 09 42.7 &24.478$\pm$0.036&1.157 & " &0.58$^{+0.16}_{-0.15}$ &1.61\\
10.3 &00 26 33.65 &17 09 42.6 &26.618$\pm$0.082& 1.042& " &0.85$^{+0.31}_{-0.26}$ &0.42\\
11.1 &00 26 34.31 &17 09 56.6 &26.927$\pm$0.088& ---& 2.5$\pm$0.5& 2.83$^{+0.38}_{-2.67}$ &0.25\\
11.2 &00 26 37.34 &17 09 08.6 &24.773$\pm$0.033&-0.407 & " &2.80 $\pm$ 0.37 &0.06\\
11.3 &00 26 38.05 &17 09 18.1 &25.326$\pm$0.042&0.026 & " &2.74$^{+0.37}_{-2.49}$ &1.01\\
11.4 &00 26 38.45 &17 09 37.8 &26.025$\pm$0.058&-0.452 & " &3.22$^{+0.41}_{-2.99}$ &1.28\\
\hline\hline
\end{tabular}
\end{center}
\end{table*}

\begin{figure}
\centering{
  \includegraphics[trim = 3cm 8cm 2cm 8cm, clip, width=8cm]{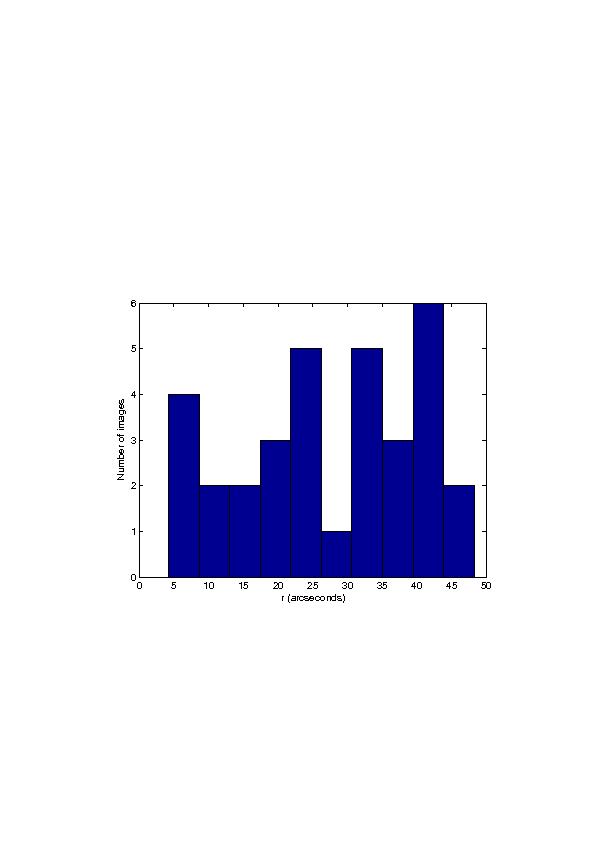}
  \caption{Radial distances from the cluster centre of all multiply-lensed images.}\label{fig:hist_images}}
\end{figure}

\begin{figure}
  \includegraphics[trim = 2cm 1cm 2cm 2cm, clip, width=9cm]{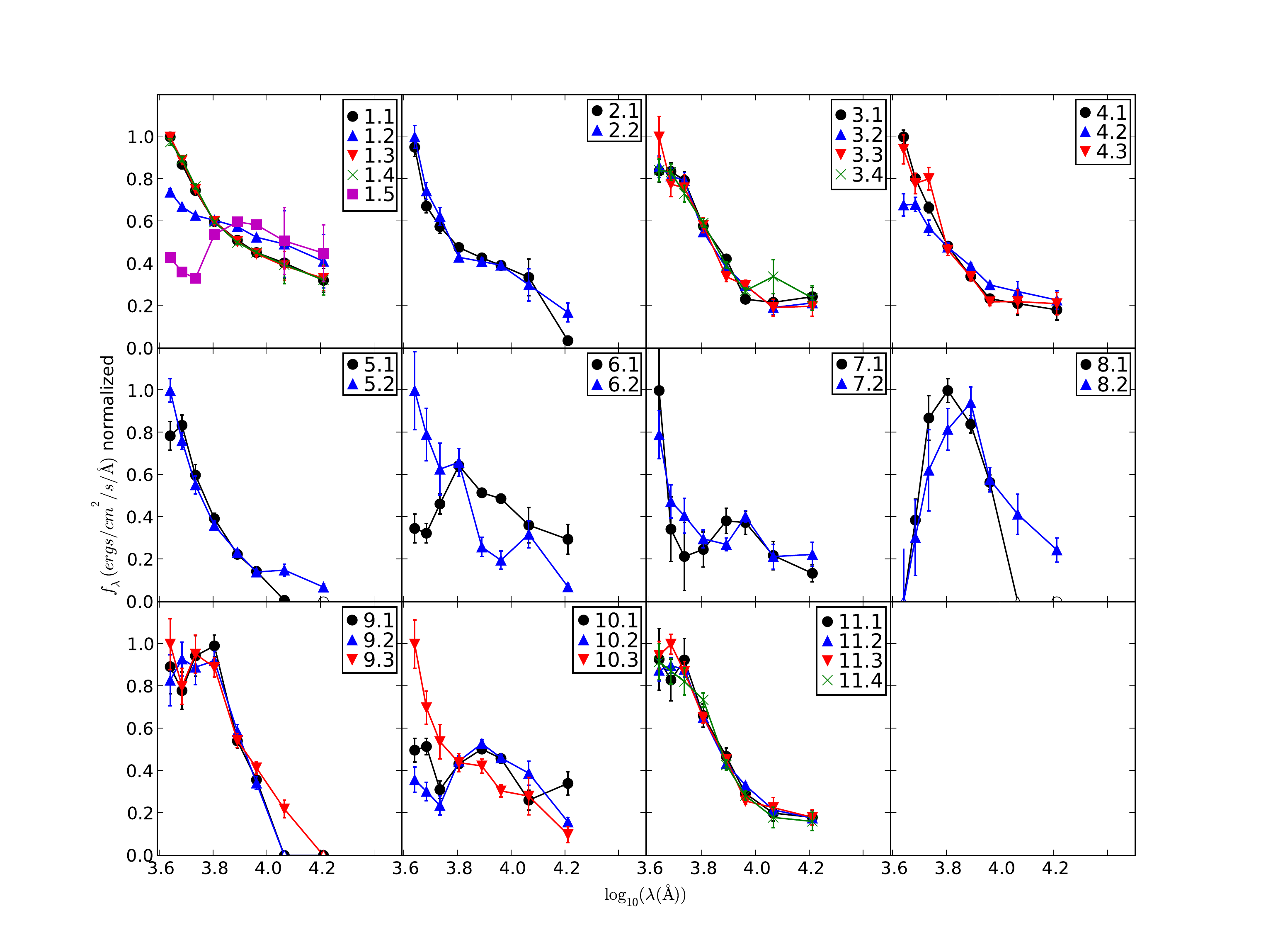} \caption{SED
  comparison of multiple image candidates of each lensed galaxy.  We
  plot $B V g\arcmin r\arcmin i\arcmin z\arcmin J H$ photometry from
  ACS and NIC3 images.  Good agreement is expected unless the light is
  contaminated by neighbouring galaxies.  We identify these cases by
  visual examination of the images.}\label{fig:seds}
\end{figure}

To summarize, all the multiply-lensed are well reproduced by the
model, with respect to their location, shape and photo-$z$. Note that
the various images within each system have similar SEDs, where not
strongly affected by galaxy light. All images are summarised in Table
\ref{systems}.

\subsection[]{Magnification Map and Critical Curves}

\begin{figure}
\centering{
  \includegraphics[trim = 1cm 1cm 1cm 11cm, clip, width=8cm]{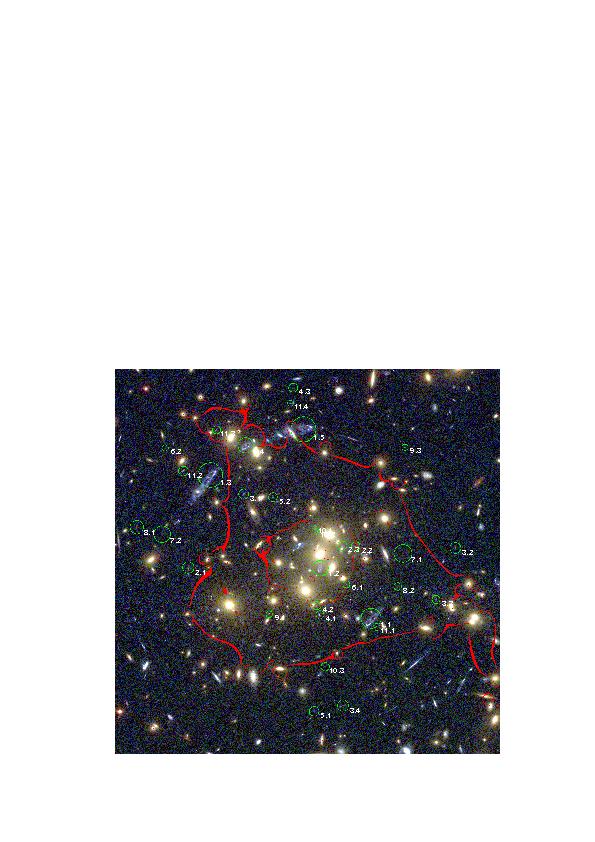}
  \caption{Magnification map (of a source at $z=1.675$) laid over ACS colour-combined image, along with the multiply-lensed images marked in green. North is right, east is up. The field of view is $\sim 100\arcsec \times100 \arcsec$.}\label{fig:MagMap}}
\end{figure}

\begin{figure}
\centering{
  \includegraphics[trim = 1cm 1cm 1cm 11cm, clip, width=8cm]{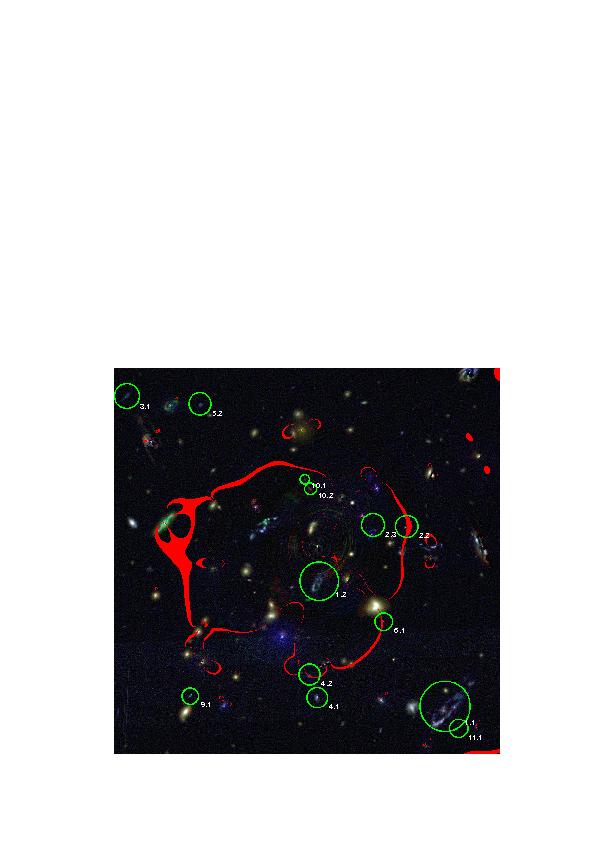}
  \caption{A galaxy subtracted image, zoomed-in on the radial critical curve and the images defining it. The radial curve in the image is for a source at $z=1.675$. North is right, east is up. The field of view is $\sim 40\arcsec \times40 \arcsec$.}\label{fig:MagMapZoom}}
\end{figure}

\begin{figure}
\centering{
  \includegraphics[trim = 3cm 8cm 2cm 10cm, clip, width=8cm]{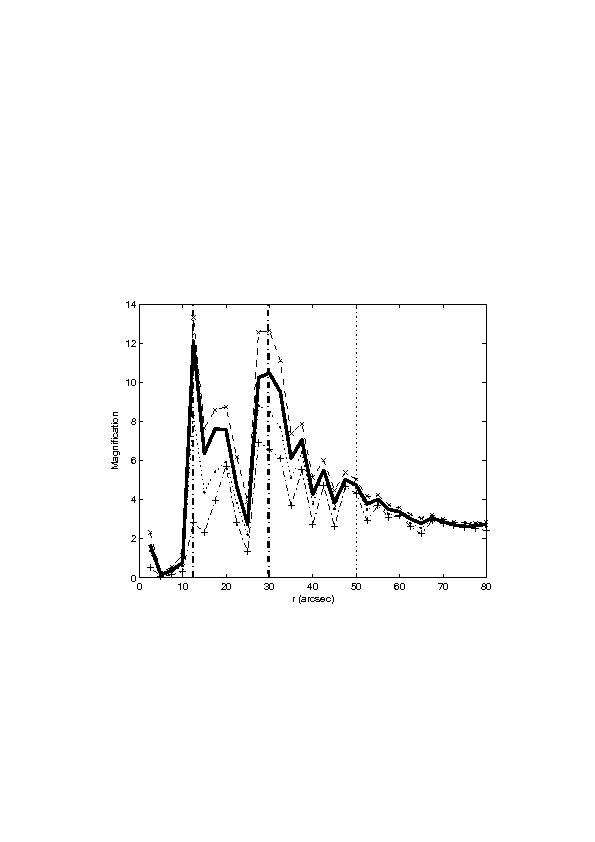}
  \caption{The resulting magnification profile. The solid thick line
  is the best model magnification-profile, where the errors are
  represented by the dashed lines denoted as in Figure
  \ref{fig:Kappa}. The dashed vertical lines are the radial and
  tangential critical radii (as determined by eye), and the $50\arcsec$ line which indicates the range within which multiple images were found.}\label{fig:Magprofile}}
\end{figure}

The magnification value at each point determines how magnified
a lensed image would be were it to appear at that point. We
compare the relative magnitudes of various images of the same system
to the same relative-magnification predicted values. Generally, this
comparison would be mostly efficient away from the critical curves and
cluster galaxies, where the magnification diverges or is highly
perturbed. We use this information as a consistency check as shown in
Figure \ref{fig:relmag} and in Table \ref{systems}. For this purpose
we use here the $i'_{775}$ magnitudes, since it better constrains the
magnitudes of higher-$z$ galaxies. Other important aspects of the
magnification map are the locations of the tangential and radial
critical curves, were the latter to exist. The location of the tangential critical curve determines
(for a certain source redshift) where highly-magnified and
significantly stretched images would form. The location of the radial
critical curve determines where and whether radial
arcs, pointing towards the cluster centre, should form. The tangential and radial critical curves,
laid on the colour cluster image are shown in Figure \ref{fig:MagMap}
(along with the multiply-lensed systems). Also, we show in a separate
Figure (\ref{fig:MagMapZoom}), the radial images around the radial
critical curve. In addition, the magnification profiles of the best-fitting models are plotted in
Figure \ref{fig:Magprofile} where we also plot the radii of the
critical curves.

\subsection{The Mass Profile and Comparison with Weak Lensing}

The mass profiles of the
acceptable best-fitting models (Figure \ref{fig:Kappa}) are very similar to each other within the errors, with a mean slope of $d\log M/d\log r\simeq
-0.55\pm 0.06$ (see also Figure \ref{fig:WL}). These profiles are effectively limited
to a maximum radius of approximately 50\arcsec, or $\sim$250
kpc/$h_{70}$, corresponding to the outer limit of our multiply-lensed
galaxies, which is approximately twice the Einstein radius. The slope we measure is similar to that obtained by Broadhurst et al. (2005a) based on a 106 multiple images in A1689, and for A1703 (Saha \& Read 2009, Figure 5) based on 42 images identified by Limousin et al. (2008).

In the very inner region we do not detect arcs at a radius below
$\sim5\arcsec$, setting an effective lower limit of $\simeq$27
kpc/$h_{70}$. In defining a meaningful radial profile we must have a
good idea of the centre of mass and also the effect of any
substructure. We take as our preferred estimate of the centre of mass to be the
centre of the inner radial critical curve which is found to be fairly
circular. This centre is close, within 3\arcsec, (16 kpc/$h_{70}$) but not
coincident with the brightest cD galaxy. There is no reason to expect a coincidence between the centre of mass of a cluster and a
cD galaxy, an object which presumably follows a complex inner orbit
about a dynamically evolving centre of mass. In some cases significant
velocity differences are found between the cD galaxy in a cluster and
the systemic velocity of a cluster determined from many cluster
members. A small but significant difference of this sort was also seen
in the detailed mass map recovered for A1689 where many central images
help to constrain well the mass distribution. SL models are
often built with fewer multiple images and often the location of the
cD galaxy is taken to be the projected centre of mass of the cluster
for want of better data.

\begin{figure}
\centering{
  \includegraphics[trim = 1cm 1cm 1cm 1cm, clip, width=8cm]{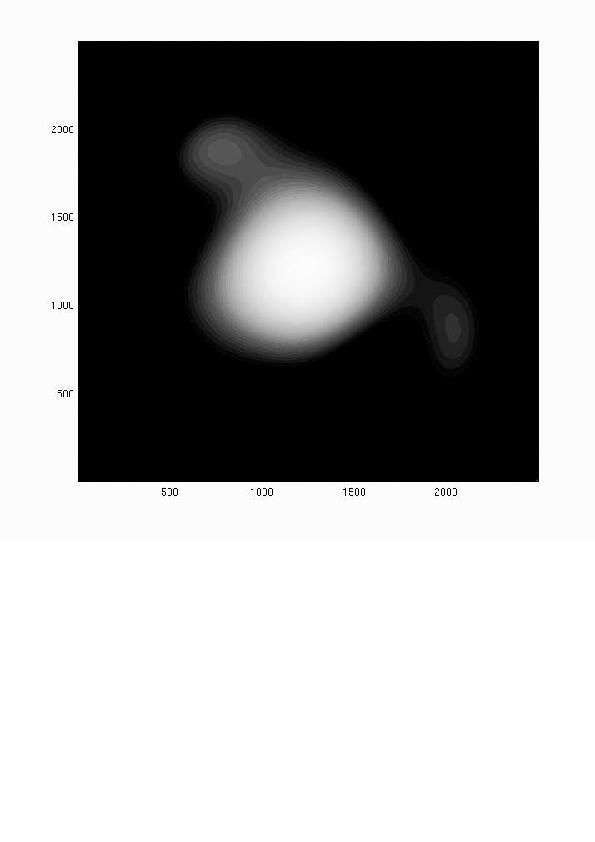}
  \vspace{-4cm}
  \caption{The resulting smooth DM distribution. Note that it is fairly circular with two small clumps. North is right, east is up. Axes are in pixels (where the pixel scale is 0.05 \arcsec/pixel).}\label{fig:DMMap}}
\end{figure}
\begin{figure}
\centering{
  \includegraphics[trim = 1cm 2cm 1cm 1cm, clip, width=8cm]{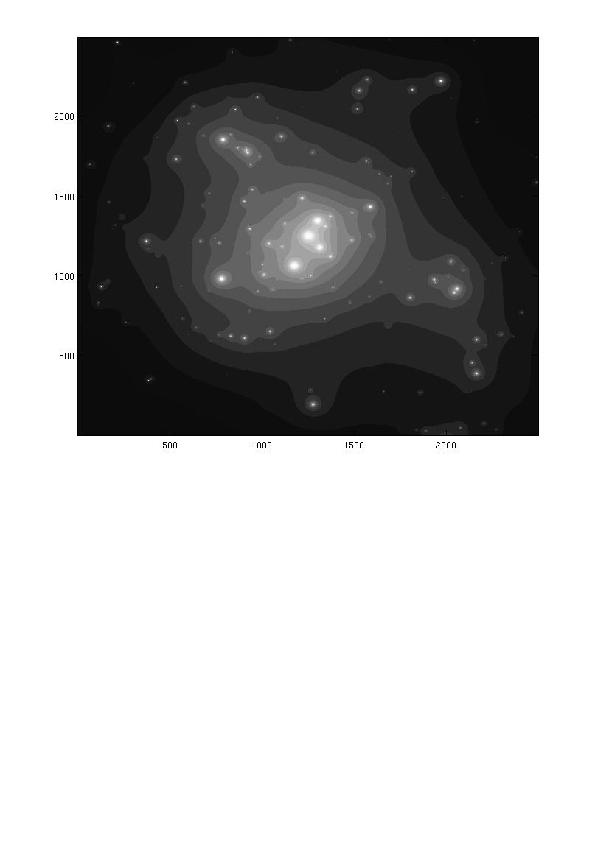}
  \vspace{-4cm}
  \caption{The total (galaxies+DM) final mass distribution. North is right, east is up. Axes are in pixels.}\label{fig:MassMap}}
\end{figure}

In 2D the mass distribution shows two small substructures containing
less than $\sim$ 10\% of the total mass within the SL area. Overall
the central mass distribution is fairly symmetric and centrally
concentrated similar to the distribution of bright cluster members
though not strictly in proportion, as discussed in section \S
\ref{ss:DM}. We may compare this radial mass profile with that
recently obtained from a new weak lensing analysis by Umetsu et
al. (2009, in prep). This is based on deep multi-colour Subaru imaging
in B'R'Z' and a two colour selection of red and blue background galaxies,
similar to that employed by Medezinski et al. (2009, in prep) for a
sample of other deeply-imaged clusters observed with Subaru. Good
agreement is found between the SL mass profile and the WL as
reconstructed from the WL measurements in the region where the
profiles overlap, as shown in figure \ref{fig:WL}.

This agreement is very reassuring, particularly because the WL
determination is model-independent. Also this demonstrates that there is no mismatch when systematic
effects, in particular the selection of background galaxies, are
properly dealt with. Dilution of the WL signal by cluster members and
unlensed foreground galaxies has been identified as a longstanding
widespread problem (Broadhurst et al. 2005b), responsible for the
often much smaller Einstein radius predicted for models fitted to WL
data than the observed Einstein radius.

If we fit an NFW model to the joint weak and strong lensing profile we obtain a
reasonable fit with a concentration of $c_{\rm vir}=8.6\pm1.6$ (Umetsu
et al. 2009, in prep.). The inner profile is somewhat shallower
than the canonical NFW profile best-fitting model to the overall data
- and similar to that obtained for A1689 by Broadhurst et al.
(2005a). A significant substructure is found in the WL maps at a
radius of about $3.5\arcmin$ from the cluster centre, in good
agreement with the location of a secondary sub-structure found by
Kneib et al. (2003), based on independent mosaic HST imaging. The
effect of this substructure, although modest, is significant in the
derivation of the overall concentration and is treated in full detail
by Umetsu et al. (2009, in prep.). Here we stress only the good
empirical agreement between the weak and strong lensing in the region
of overlap which we take as independent support for our SL
solution.

\begin{figure}
\centering{
  \includegraphics[trim = 6cm 2.5cm 1cm 5cm, clip, width=7cm,angle=270]{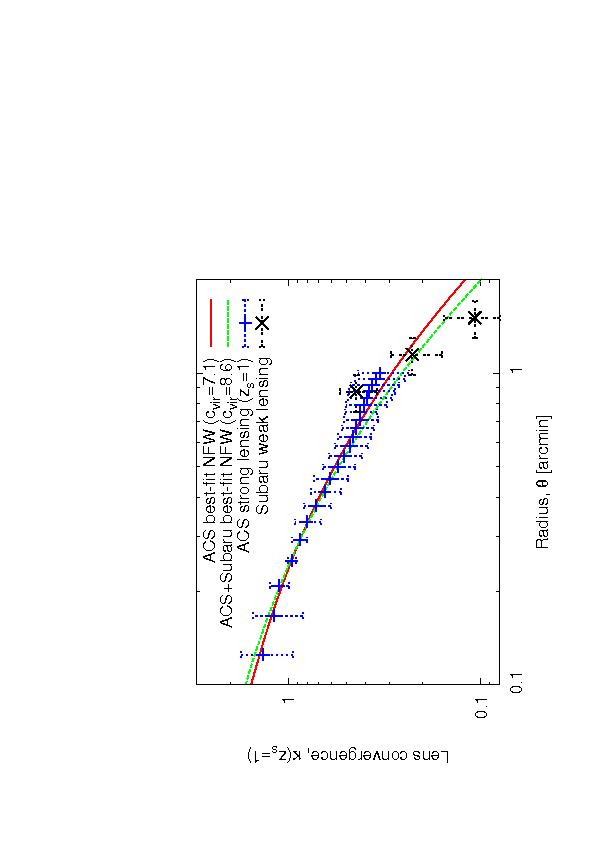}
  \caption{Comparison of weak and strong lensing determinations of surface mass densities. Good consistency is found here between these very different, independent measurements. The best fitting NFW profile is indicated, for SL alone (red), and in combination with WL (green), see Umetsu et al. (2009, in perp).}\label{fig:WL}}
\end{figure}

\begin{figure}
\centering{
  \includegraphics[trim = 2.2cm 1cm 2cm 14cm, clip, width=8cm]{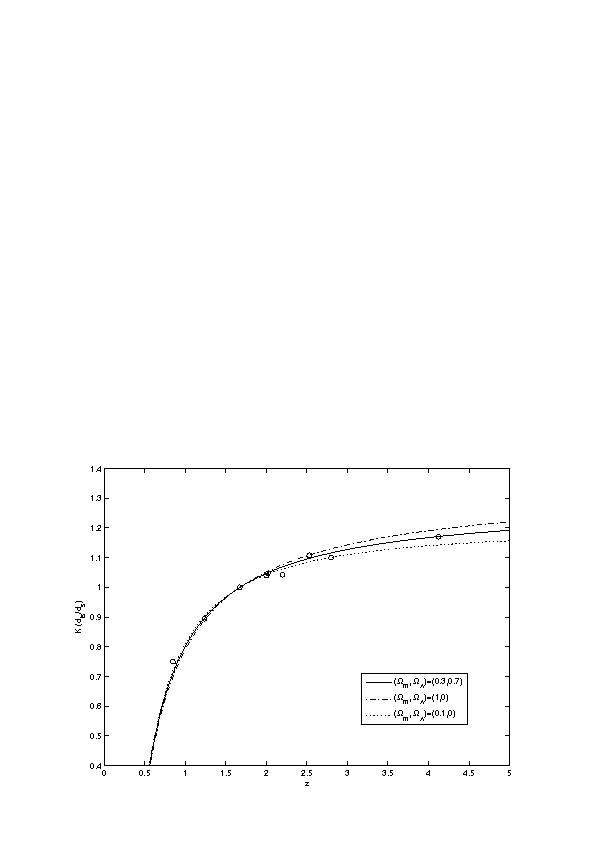} \caption{Growth of the
  scaling factor $K_{q}$ as a function of redshift, normalised so $K_{q}$=1 at
  $z=$1.675. Plotted lines are the expected ratio from the chosen
  specified cosmology. The points are the multiple-image systems. The data
  follow very well the relation predicted by the standard cosmology.}\label{fig:cosmo}}
\end{figure}
\begin{figure}
\centering{
  \includegraphics[trim = 3cm 8cm 2cm 8cm, clip, width=8cm]{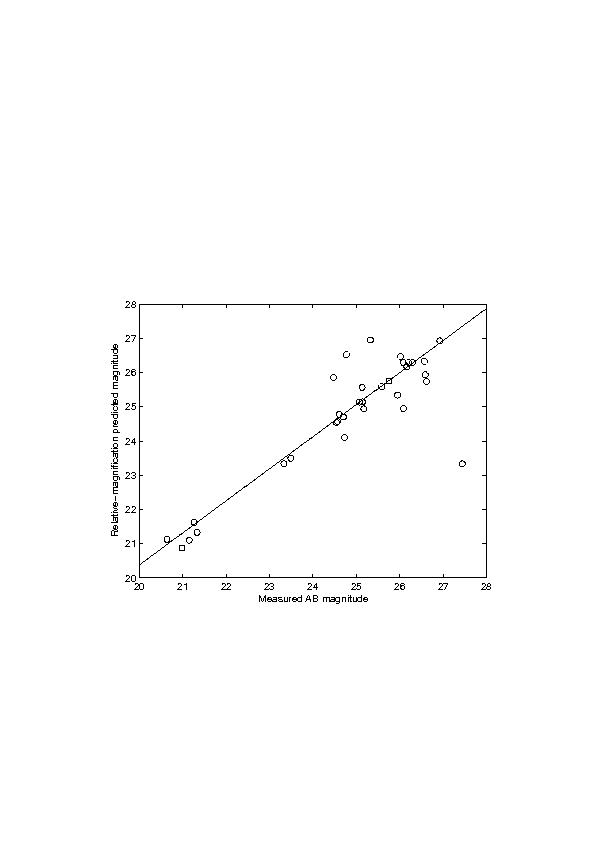}
\caption{Relation between the measured AB magnitudes and the model-predicted
  magnitudes, as derived from the relative magnifications as explained
  in \S \ref{sss:criteria3}. As can be seen, these are closely proportional.}\label{fig:relmag}}
\end{figure}

\section[]{Summary and conclusions}

We have presented a simple and accurate strong-lensing model for the
galaxy cluster Cl0024+1654, based on ACS/NIC3/GTO images. This modelling has identified many multiply-lensed systems, which
in turn were incorporated into the model to improve the fit, mainly
by the use of their photo-$z$ measurements. We have found that for the
purpose of identifying multiple images, a wide range of profile slopes
can be tolerated. This is an inherent lensing degeneracy that can be
made use of, in the sense that those parameters controlling the overall profile slope can be initially set to reasonable values, enabling detection of multiply-lensed systems using only 4 free parameters, which are fully constrained by known systems.
However, the model-predicted redshift for each system is sensitive to
the profile slope and so we have made use of our relatively accurately-determined photo-$z$ measurements to constrain the slope by minimising the
parameters controlling the profile slope with respect to
the standard cosmological distance-redshift relation.

The presented SL modelling-method uses only 6 free parameters,
which is significantly less than other competing methods, in particular
the model-independent approaches where the lens plane is described by a
set of large pixels or with orthogonal functions etc. One relative
disadvantage of our method is manifested in the reproduction accuracy
of the multiply-lensed systems, with respect to the image-plane RMS
and internal shape. For example, here the image-plane RMS is typically
$\sim2.5\arcsec$ $per~image$, whereas Broadhurst et al. (2005a) achieved
an RMS of $3.2\arcsec$ per image for A1689, and later Halkola, Seitz,
\& Pannella (2006) reported an RMS of $2.7\arcsec$ per image for that
cluster. Quantifying these results relative to the size of the Einstein radius, Broadhurst et al. (2005a) and Halkola, Seitz,
\& Pannella (2006) obtained a $\sim$20\% better image-plane RMS than achieved here. Naturally, a higher number of parameters would supply a more accurate
solution, however, the efficiency of a model decreases extremely
rapidly as more parameters are added to the minimisation procedure,
and the confidence in the model decreases as more arbitrary non-physical
parameters are added, as is often the case in other methods. We have shown that the presented method, with only 6 free parameters built on simple physical considerations, does a very good job in finding new multiply-lensed systems and in
constraining the deflection field and accordingly the mass distribution and profile.

A smooth low-order fit is found to be a good representation of the overall mass
distribution of the cluster, with a small contribution required from
the observed cluster member galaxies. With this model we have
accurately reproduced the well known 5-image system in this cluster
field, and confirmed the tentative 2 image system identified in
earlier WFPC-2 based modelling (Broadhurst et al. 2000), finding an
additional third image associated with this source. In addition we
identify 9 other multiple-images systems, bringing the total known for
this cluster to 33 multiply-lensed images, spread fairly evenly all
over the central area, $r\leqslant50\arcsec$. We stress that our multiple images are accurately reproduced by our
model and not simply eyeball candidates requiring redshift
verification.

Our best fitting model minimises the image-plane RMS of the
reconstructed image location compared with the observed positions. We
also find this model satisfies three other completely independent
criteria. Firstly the best fit solution recovers the expected
cosmological relation of $d_{ls}/d_{s}$ versus redshift for all sources with reliable photometric redshifts. We also find that the
predicted and derived relative magnifications of the multiple images
are also well reproduced (see Figure \ref{fig:relmag}). In addition, we find good agreement between our SL-based mass profile
and the mass profile recovered from WL in the region of overlap of
these profiles, implying an NFW profile with a concentration of $c_{\rm vir}=8.6\pm1.6$. These independent consistency checks reassure us that
the derived mass distribution is a reliable representation of the
central mass distribution. The success of this simple and minimalistic
method we have applied here motivates an analysis of larger
statistical well-defined samples of massive relaxed clusters, which
should be helpful in constraining in a new way the cosmological
curvature, and shed more light on the nature of DM by comparison with
the increasingly accurate predictions for the properties of DM
dominating massive galaxy clusters.

\section*{acknowledgments}
Zitrin A. and Broadhurst T. gratefully thank Paul Ho for the hospitality of ASIAA, where much of this work was accomplished.
ACS was developed under NASA contract NAS 5-32865, and this research has been supported by
NASA grant NAG5-7697. Results are partially based on observations made with the NASA/ESA Hubble Space Telescope, obtained from the data archive at the Space Telescope Science Institute. STScI is operated by the Association of Universities for Research in Astronomy, Inc. under NASA contract NAS 5-26555. Zitrin A. thanks Eran Ofek for his publicly available Matlab scripts and Assaf Horesh for useful programming advise.
This work is in part supported by the Israeli Science Foundation and the National Science Council of Taiwan under the
grant NSC95-2112-M-001-074-MY2. Part of this work is based on data collected at the Subaru Telescope, which is operated by the National Astronomical Society of Japan. Ascaso B. is partially supported by NASA grant NNG05GD32G.

\bsp
\label{lastpage}

\end{document}